\documentclass{article}
\usepackage[utf8]{inputenc}

\usepackage{import}
\usepackage[utf8]{inputenc}
\usepackage[british]{babel}
\usepackage{csquotes}
\usepackage{textcomp}

\usepackage{amssymb, amsmath, amsthm, amsfonts, mathrsfs}
\usepackage{amsthm}
\usepackage{slashed}

\usepackage{tikz-cd}

\usepackage[shortlabels]{enumitem}
\usepackage{listings}

\usepackage[backend=biber,style=numeric-comp,sorting=none,maxbibnames=9]{biblatex}
\usepackage{hyperref}
\hypersetup{
    colorlinks = true,
    linktocpage = false,
    linkcolor = blue,
    citecolor = red,
    urlcolor = blue
}
\usepackage{bibentry}

\usepackage{appendix}
\usepackage[a4paper,margin=1.5in]{geometry}
\usepackage{setspace}

\usepackage{subfigure}

\usepackage{anyfontsize}

\usepackage[notrig]{physics}

\definecolor{cambridgeblue}{rgb}{0.64, 0.76, 0.68}
\definecolor{lapislazuli}{rgb}{0.15, 0.38, 0.61}
\definecolor{awesome}{rgb}{1.0, 0.13, 0.32}
\definecolor{aureolin}{rgb}{0.99, 0.93, 0.0}
\definecolor{almond}{rgb}{0.94, 0.87, 0.8}
\definecolor{antiquewhite}{rgb}{0.98, 0.92, 0.84}
\definecolor{oxfordblue}{rgb}{0.0, 0.13, 0.28}

\graphicspath{ {Images/} }

\begin{document}
\begin{titlepage}

\vfill

\begin{center}
	\baselineskip=16pt  
	
{\Large \bf  Asymptotics in an Asymptotic CFT}

	\vskip 1cm
	{\large \bf    Lucas Schepers$^{a}$\footnote{\texttt{lucas.schepers@swansea.ac.uk}}   and    Daniel C. Thompson$^{a,b}$\footnote{\texttt{daniel.c.thompson@swansea.ac.uk}} }
	\vskip .6cm
	{\it  
			$^a$ Department of Physics, Swansea University, \\ Swansea SA2 8PP, United Kingdom \\ \ \\
                $^b$ Theoretische Natuurkunde, Vrije Universiteit Brussel,\\ \& The International Solvay Institutes, Pleinlaan 2, B-1050 Brussels, Belgium \\ \ \\
			}
	\vskip 2cm
\end{center}

\begin{abstract}
\noindent
In this work we illustrate the resurgent structure of the $\lambda$-deformation;  a two-dimensional integrable quantum field theory that has an RG flow with an $SU(N)_k$ Wess-Zumino-Witten conformal fixed point in the UV.  To do so we use modern matched asymptotic techniques applied to the thermodynamic Bethe ansatz formulation  to compute the free energy to 38 perturbative orders in an expansion of large applied chemical potential.   We find numerical evidence for factorial asymptotic behaviour with both alternating and non-alternating character which we match to an analytic expression.   A  curiosity of the system is that it exhibits the Cheshire Cat phenomenon with the leading non-alternating factorial growth vanishing when $k$ divides $N$.    The ambiguities associated to Borel resummation of this series are suggestive of non-perturbative contributions.  This is verified with an analytic study of the TBA system demonstrating a cancellation between perturbative and non-perturbative ambiguities. 
\end{abstract}

\vfill

\setcounter{footnote}{0}
\end{titlepage}




\section{Introduction}

A complete understanding of the strong coupling dynamics of four dimensional asymptotically free (AF) non-supersymmetric  gauge theory, i.e. QCD,  remains elusive.  To gain a foothold we may  turn to simplified toy models.  One strategy is to reduce the dimensionality of the problem considering instead two dimensional quantum field theories with similar RG behaviour.  In the special case of integrable QFTs,  an infinite set of symmetries  completely determine the exact S-matrix \cite{zamolodchikov1977exact} providing a powerful  toolkit that can be used to tackle non-perturbative questions.  An early success of this approach was the calculation of the exact ratio of the mass gap  to cut-off of AF integrable  QFTs  theories \cite{hasenfratz1990exact1,hasenfratz1990exact2,Forgacs:1991rs,Forgacs:1991ru}.

More recently, techniques in integrable models have been used to elucidate even deeper questions of the nature of perturbation theory.  Typically, perturbation theory is asymptotic in nature with perturbative coefficients growing factorially.  The programme of resurgence asserts that this breakdown of convergence signals the need to include non-perturbative physics.  Even more strongly, ambiguities inherent in resummations of asymptotic perturbative expansions should be cancelled by compensating ambiguities in a non-perturbative sector.  For a modern overview of resurgence from a physics view point see e.g. \cite{aniceto2018primer}.  Again, integrable two-dimensional models provide an ideal test bed for resurgence.  

In semi-classical approaches \cite{ dunne2012resurgence,Cherman:2013yfa,cherman2015decoding,Dunne:2015ywa,demulder2016resurgence,Schepers:2020ehn}, an adiabatic compactification of two-dimension models is used to obtain a quantum mechanics which can be probed to large perturbative orders. In these cases, two-dimensional finite Euclidean action  configurations,  known as unitons\footnote{Unlike instantons, unitons are not topologically protected.} are shown to precisely resolve the semi-classical ambiguities of the perturbative sector.   Whilst intriguing, such approaches intrinsically disregard degrees of freedom in compactification restricting to the lowest KK sector. Alongside this features of renormalisation group are disregarded in the truncation to Quantum Mechanics.  Given these limitations, one thus prompted to ask if the resurgence paradigm can be established in a fully two-dimensional setting. 

 A breakthrough was the work of Volin \cite{volin2010mass,volin2011quantum}, recently refined in \cite{Marino:2019eym, Marino:2019wra} (see also the recent papers \cite{Marino:2019fvu, Marino:2021six, Bajnok:2021zjm, Bajnok:2022ucr, Bajnok:2022rtu,Bajnok:2021dri,Abbott:2020qnl,Abbott:2020mba, Bajnok:2022xgx}), in addressing the Thermodynamic Bethe Ansatz (TBA) system that determines free energy in a large chemical potential.   By comparing two scaling limits, it is possible to reduce the complicated integral TBA equation to a (complicated) set of algebraic equations that fix unknown coefficients in an ansatz for a perturbative expansion (in the chemical potential or other more refined coupling).    This allows access to sufficient order in perturbation theory to reveal  factorial divergence of perturbative coefficients.  

Although we cannot identify an instanton or other semi-classical non-perturbative saddle in the TBA approach, we can find a matching ambiguity using a different method. \cite{Marino:2021dzn} showed that it is possible to solve the TBA equations using a transseries. A critical step in their solution is an arbitrary choice of branch cut which introduces an ambiguity of the transseries. Although this approach, due to its computational difficulty, cannot be executed to large orders, it does exhibit a non-perturbative ambiguity that matches the ambiguity of the large-order behaviour found in the perturbative sector.

 In this note we will adopt this toolbox to study the resurgent structure of a theory that exhibits a different renormalisation group dynamic. We consider a theory in which the UV is not Gaussian but instead is described by a non-trivial interacting conformal fixed point.   The theory we will consider, known as the $\lambda$-model \cite{Sfetsos:2013wia, Sfetsos:2014jfa}, is realised as a flow away from an $SU(N)_k$  Wess-Zumino-Witten (WZW) model driven at leading order by a certain current-current bilinear. The IR of the theory is the principal chiral model, expressed in non-Abelian T-dual coordinates, and accordingly is gapped.  Whilst this marginally relevant deformation breaks conformality and the full affine symmetry of the WZW current algebra, it does preserve an infinite symmetry associated to integrablity.  At the quantum level the exact S-matrix is known (based on symmetry grounds pre-dating the Lagrangian description) \cite{Evans:1994hi, Evans:1995dn} and has been shown to match the $\lambda$-model Lagrangian using a light cone lattice discretisation and Quantum Inverse Scattering \cite{appadu2017quantum}.

 The goal of this note is to match the perturbative ambiguity to that of the transseries the new context of a $\lambda$-model. The outline is as follows:  Section \ref{sec:lambda} provides a more in-depth discussion of the $\lambda$-model as we consider its RG flow in more detail and present its exact S-matrix. In Section \ref{sec:tba}, we review the recent techniques to perturbatively solve TBA equation \cite{volin2010mass, volin2011quantum, Marino:2019eym, Marino:2019wra} that determine free energy. Introducing a special coupling $\gamma$ in Section \ref{sec:PertTBA} results in a clean (i.e. log-free) series for the $\lambda$-model. We analyse its asymptotic behaviour in Section \ref{sec:Asymp} and compute the leading ambiguity. This ambiguity is matched by a transseries calculation in Section \ref{sec:transseries}. A particularly eye-catching result is that the leading UV ambiguity disappears when $N$ divides $k$. We wrap up with ideas for future research in Section \ref{sec:outlook}.

\section{The \texorpdfstring{$\lambda$}{lambda}-Model } \label{sec:lambda}

In this section we outline the salient properties of the two-dimensional integrable QFT that we are considering: the $\lambda$-deformed model.   
Classically, the $\lambda$-model provides a Lagrangian interpolation between the conformal Wess-Zumino-Witten (WZW) model for a Lie-group $G$ \cite{witten1994non} and the principal chiral model (PCM) (written in non-Abelian T-dual variables).  Remarkably this theory is integrable for all values of the eponymous interpolating parameter $\lambda$  related to the level, $k$, of WZW and the radius, $r$, of the PCM by
\begin{equation} \label{eq:lambda}
    \lambda = \frac{k}{k + r^2}\,.
\end{equation}

 Whilst  the $\lambda$-model for the restricted case of $G= SU(2)$  was first proposed long ago  \cite{Balog:1993es, Sfetsos:1994vz, Evans:1994hi},  the pioneering work of Sfetsos   \cite{Sfetsos:2013wia} in constructing the general theory has prompted extensive recent development (for reviews see \cite{Thompson:2019ipl, Hoare:2021dix}).   The $\lambda$-model has been extended to $\mathbb{Z}_2$ graded symmetric spaces \cite{ Sfetsos:2013wia,Hollowood:2014rla} where it constitutes an interpolation between a $G/H$ gauged WZW (representing the coset CFT) and the (non-abelian T-dual of) the principal chiral model on $G/H$ and even to $\mathbb{Z}_4$ graded super-cosets relevant to the construction of the $AdS_5 \times S^5$ superstring \cite{hollowood2014integrable} underpinned by an elegant quantum group at root-of-unity symmetry structure.  In this string theory context, $\lambda$-deformation is in fact marginal, and the world sheet theory can be viewed as a $\sigma$-model in some target space super-gravity background  \cite{Sfetsos:2014cea,Demulder:2015lva,Borsato:2016ose}.  Here however we will be considering the simpler bosonic case for which the $\lambda$-deformation does not define a CFT but rather a relevant RG flow from a WZW fixed point in the UV to  the dualised PCM the IR \cite{Tseytlin:1993hm, Sfetsos:2014jfa, Appadu:2015nfa}.   A series of papers \cite{Klimcik:2016rov, Hoare:2015gda, sfetsos2015generalised, Vicedo:2015pna, Klimcik:2015gba} have shown how the $\lambda$-model is actually part of a wide tapestry of integrable deformed models linked by (analytically continued) Poisson-Lie T-duality transformations. 

 \subsection{Lagrangian Construction}

First we sketch the construction of the non-abelian T-dual of the PCM using the Buscher procedure \cite{Buscher:1987qj} as it informs the construction of the  $\lambda$-model. We start with the action of the PCM for a group valued field $\tilde{g}$ \footnote{We use light cone coordinates $\sigma_\pm=\frac{1}{2}(t\pm x)$. Derivatives with respect to light cone coordinates are denoted by $\partial_\pm$.}
\begin{equation} 
    S_{\text{PCM}}[\tilde{g}] = - \frac{r^2}{4 \pi} \int d^2\sigma\Tr \left( \,\tilde{g}^{-1} \partial_+ \tilde{g}  \tilde{g}^{-1} \partial_- \tilde{g}\right)\,,
\end{equation}
and downgrade the left symmetry $\tilde{g} \rightarrow h^{-1} \tilde{g}$ to a gauge symmetry by introducing a gauge connection transforming as $A \rightarrow h^{-1} d h +  h^{-1} A h$ and  replacing derivatives to covariant derivatives $d \rightarrow D = d + A$.  This yields the gauged PCM action which we denote as $S_{\text{gPCM}}[\tilde{g}, A]$.  To ensure that the gauged theory is actually equivalent to the ungauged theory (at least in trivial topology which we assume throughout) we enforce that the connection is flat  (i.e. the gauge field is pure gauge). This is implemented by introducing a Lagrange-multiplier term, $-\Tr ( \nu F_{+-})$, to the Lagrangian. Integrating out the field $\nu$ enforces that the field strength $F_{+-}$ vanishes and we recover the original PCM after gauge-fixing $\tilde{g}=1$. However, if instead we integrate out the gauge fields $A$, after gauge fixing $\tilde{g}=1$, we obtain the non-abelian T-dual model in which the field $\nu$ becomes the fundamental field. 

The construction of the $\lambda$-model by Sfetsos \cite{Sfetsos:2013wia} is achieved through a modification of this Buscher procedure.  Instead of adding a Lagrange multiplier term, we add a gauged WZW term. Recall that the WZW model is given by
\begin{equation}
\begin{split}
    S_{\text{WZW},k}[g] &=  - \frac{k}{2 \pi} \int_\Sigma d^2\sigma \, \Tr \left( g^{-1} \partial_\mu g g^{-1} \partial^\mu g\right) - \frac{ik}{6 \pi} \int_{M_3} \Tr \left( g^{-1} d g \right)^3 \,   ,
\end{split}
\end{equation}
in which  $g$ is extended to a 3-manifold $M_3$ with boundary $\partial(M_3) = \Sigma$. Standard arguments \cite{witten1994non} ensure that the path integral is well-defined (independent of choice of extension) provided that $k$ is appropriately quantised, and in particular for $G= SU(N)$ which we assume henceforth, $k \in \mathbb{Z} $.  In this sector we gauge the diagonal symmetry $g \rightarrow h^{-1} g h$ leading to a gauged WZW model action $S_{\text{gWZW},k}[g, A]$ \cite{Gawedzki:1988hq, Witten:1991mm}.

To construct the $\lambda$-model we combine a gauged PCM and a gauged WZW model:
\begin{equation}
    S_{\lambda,k}[g,\tilde{g}, A] = S_{\text{gPCM}}[\tilde{g}, A] + S_{\text{gWZW}}[g, A]\, . 
\end{equation}
Notice that the two models are coupled through the fact that they are gauged by the \textit{same gauge field}. The Sfetsos procedure is concluded by gauge fixing $\tilde{g}=1$ and integrating out the gauge field $A$ using its on-shell value
\begin{equation} \label{eq:Aonshell}
    A_+ = \lambda (1- \lambda \text{Ad}_g)^{-1} R_+ \,, \qquad A_- = - \lambda (1- \lambda \text{Ad}_{g^{-1}})^{-1}L_- \,,
\end{equation}
where we defined $R_\pm = \partial_\pm g g^{-1}$ and $L_\pm = g^{-1} \partial_\pm g$ and the adjoint action $\text{Ad}_g X = g X  g^{-1}$.  
Integrating out the gauge field then yields the action 
\begin{equation}
    S_{\lambda,k}[g] = S_{\text{WZW},k}[g] + \frac{k\lambda}{\pi}\int_\Sigma d^2\sigma\,  \Tr \left(  R_+ (1- \lambda \text{Ad}_g)^{-1} L_- \right) \,. 
\end{equation}
 Though not vital for what follows we note that the equation of motion can be understood as a zero-curvature condition on the Lax connection  \cite{Sfetsos:2013wia, Hollowood:2015dpa} 
 \begin{equation}
    \mathcal{L}_\pm(z) = - \frac{2}{1+\lambda}\frac{A_\pm}{1 \mp z}\,,
\end{equation}
in which  $A_\pm$ are evaluated with the on-shell values  eq. \eqref{eq:Aonshell} and  $z\in \mathbb{C}$ is a spectral parameter. 
This is the starting point of establishing the classical integrability of the theory.  Further to this one requires {\em strong integrability} i.e. that the conserved charges built from the monodromy of this Lax are in involution.  This is ensured provided that the Poisson algebra of the spatial component of the Lax has a particular r-s Maillet form as was demonstrated for the $\lambda$-model, and its generalisations, in \cite{Itsios:2014vfa,Vicedo:2015pna,Georgiou:2019plp}.

 \subsection{Renormalisation} 
The parameter $\lambda$ given by eq. \eqref{eq:lambda} varies from $0$ to $1$ and we shall now discuss what happens in each of those limits.   At a quantum level  the parameter $\lambda$ undergoes an RG flow \cite{Tseytlin:1993hm, Appadu:2015nfa, Sfetsos:2014jfa}  given by (to all orders in $\lambda$ and  leading in $\frac{1}{k}$)  
\begin{equation} \label{eq:RGlambda}
    \mu \frac{d \lambda}{d \mu}= \beta(\lambda) = - \frac{2 N}{k} \left( \frac{\lambda}{1+\lambda}\right)^2  =  - \beta_1 \lambda^2 - \beta_2 \lambda^3+ \mathcal{O}(\lambda^4)   \,.
\end{equation} 
The leading order behaviour, which shall be relevant later,  is given by   
\begin{equation}
    \beta_1 = \frac{2 N}{k}\,, \qquad \beta_2 = - \frac{4N }{k}\,.
\end{equation}

There is an evident UV fixed point at $\lambda =0$, corresponding to the undeformed WZW model.  In the vicinity of this $\lambda\approx 0$, or $k \ll r^2$, we obtain a current-current deformation of the WZW model:
\begin{equation}
    S_{\lambda, k}[g] = S_{\text{WZW},k}[g] + \lambda \int_\Sigma d^2\sigma \, \Tr \left(  R_+ L_-  \right) + \mathcal{O}(\lambda^2)\,,
\end{equation}
This, however,  is  {\em not} a marginal deformation (cf. marginal ones \cite{Hassan:1992gi}), but  relevant as it moves away from the WZW theory located in the UV.

To understand the IR regime as $\lambda \rightarrow 1$, we can force $k\rightarrow \infty$. If the group element is expanded as $g = 1 + \frac{i}{k} \nu^a t_a$, the action $S_{\text{gWZW},k}$ reduces to the Lagrange multiplier term $-\Tr ( \nu F_{+-})$. Thus in this limit the Sfetsos procedure  reduces to the non-Abelian T-dualisation Buscher procedure described above. Hence, in this IR limit, we recover the non-abelian T-dual of the PCM.  Further into the deep IR, one thus anticipates (as with the PCM) that the dimensionless parameter $\lambda$ is transmuted into a mass gap mediated through a cut-off $\Lambda$.

\subsection{Quantum Integrability} 
\label{sec:Qintegrability}
Not only is the theory classical integrable, it remains so at the quantum level.  The existence of higher spin conserved currents ensure that the scattering matrix of the theory factorises, and can be fully determined with the 2-to-2 particle scattering matrix the fundamental building block.  The S-matrix for the $SU(N)$ $\lambda$-model was constructed many years ago \cite{ahn1990fractional} from an algebraic perspective, and was related directly to the Lagrangian description for the $SU(2)$ case in  \cite{Evans:1994hi} by matching the free energy obtained by Lagrangian perturbation theory and by S-matrix TBA techniques.  Following the introduction of the Lagrangian description $\lambda$-model by Sfetsos \cite{Sfetsos:2013wia} for $SU(N)$  the exact S-matrix was conjectured  \cite{ Hollowood:2015dpa} for general ranks.  This conjecture was  substantiated by Appadu et al.  \cite{appadu2017quantum} in which the form of the S-matrix was `derived' by the Quantum Inverse Scattering Method (i.e. a latticed version of the theory that takes the form of a spin chain such from the QFT particle states are obtained as excitations over the ground state in a continuum limit)\footnote{This QISM is in fact rather non-trivial as the  $\delta'(\sigma) $ non-ultra-local terms in the fundamental Poisson bracket preclude a simple application of QISM. Instead what is proposed is a modification of the $\lambda$-model, that lies in the same universality class, to which QISM can be applied.  This provides a description as a spin-$k$  XXX spin chain with alternating inhomogeneities.  This idea was expanded to a two-parameter integrable $\lambda$-type model \cite{Appadu:2018ioy} realised as a spin-$k$  XXZ spin chain with alternating inhomogeneities. }.    Rather than present  the full details of the S-matrix (for which see \cite{ Hollowood:2015dpa}) we can give a schematic understanding somewhat mirroring the Sfetsos procedure.   

We start with the $SU(N)$ principal chiral model which has in particular an $SU(N)_L \times SU(N)_R$ global symmetry.   The fundamental particles are massive and transform in fundamental antisymmetric tensor representations of the global symmetry.  The scattering depends kinematically only on the rapidity difference $\theta$ of the particles\footnote{The mass shell is related to rapidity by $p_0 = m \cosh \theta$ and $p_1= m \sinh \theta$.}.   Reflecting this global symmetry, the S-matrix of these fundamental excitations has a schematic tensor form (suppressing explicit representation labels)
\begin{equation}\label{eq:PCMSmatrix}
    \mathbb{S}_{\textrm{PCM}}(\theta) = X(\theta) {\cal S}(\theta) \otimes  {\cal S}(\theta) \, , 
\end{equation}
where $X(\theta)$ is an overall scalar dressing factor to ensure  all S-matrix axioms are obeyed, and the $ {\cal S}(\theta)$ factors are separately $SU(N)$ invariant (in fact invariant under a larger Yangian symmetry).   Recalling that in the Sfetsos procedure the left acting $SU(N)_L$ symmetry was gauged, it is natural that the left hand block of the tensor product of eq. \eqref{eq:PCMSmatrix} is modified in the $\lambda$-theory and indeed this is the case with
\begin{equation}\label{eq:Lambdamatrix}
    \mathbb{S}_{\lambda}(\theta) = X_k(\theta) {\cal S}_k(\theta) \otimes  {\cal S}(\theta) \, . 
\end{equation}
Here ${\cal S}_k(\theta)$ is a block  \cite{ahn1990fractional}  that furnishes a quantum group symmetry at the $q^{2(k+N)}= 1$ root of unity  taken in Restricted-Solid-On-Solid (RSOS) picture representing the scattering of kink degrees of freedom.  

Given knowledge of the exact S-matrix, the Thermodynamic Bethe Ansatz yields a set of rather complicated coupled-integral equations can be used to determine the free-energy of the theory.  Solving these is quite formidable especially as the S-matrix is non-diagonal.  A powerful simplification is achieved by exposing the system to a chemical potential  $h$ for a $U(1)$ charge such that only certain particles condense and contribute to the ground state.   When the charge is chosen appropriately (as the one defined by a highest weight of a rank $N/2$ antisymmetric representation \cite{appadu2017quantum})   then only a single particle of maximal charge contributes and the TBA system simplifies to a single integral equation determined by the identical scattering of this particle.

In this case, the scattering ``matrix" reduces to a simple phase factor $S(\theta)$ that governs transmission and reflection. It shall prove useful in this case to define the scattering kernel of this reduced S-matrix by 
\begin{equation} \label{eq:KdLogS}
    K(\theta) = \frac{1}{2 \pi i} \frac{d}{d \theta} \log S(\theta)\,,
\end{equation}
and its Fourier transform
\begin{equation} \label{eq:KFourier}
    K(\omega) = \int_{-\infty}^\infty d\theta\, e^{i \omega \theta} K(\theta)\,.
\end{equation}
As a consequence of Hermitian analyticity on the reduced S-matrix,   both  $K(\theta)$  and its Fourier transform are symmetric functions. Explicitly we have that the relevant kernel  is  given by \cite{Appadu:2018ioy}  
\begin{equation}
   1- K(\omega) = \frac{\sinh^2(\pi \omega/2)}{\sinh(\pi \omega)\sinh( \pi \kappa \omega )} \exp(\pi \kappa \omega )\,, 
\end{equation}
where $\kappa = \frac{k}{N}$.   In what follows, it shall prove useful to write the Fourier transform of the scattering kernel as a Wiener-Hopf (WH) decomposition
\begin{equation} \label{eq:WHDecomp}
    1 - K(\omega) = \frac{1}{G_+(\omega)G_-(\omega)}\,,
\end{equation}
where $G_-(\omega) = G_+(-\omega)$, and $G_+(\omega)$ is analytic in the Upper Half Plane (UHP) and normalised such that  $G_+(2 i s) = 1 + \mathcal{O}\Big(\frac{1}{s}\Big)$.  Explicitly $G_+(\omega)$ is  given by 
\begin{equation} \label{eq:WHlambda}
\begin{split} 
    G_+(\omega) &= \sqrt{4\kappa}\frac{\Gamma(1- i \omega/2)^2}{\Gamma(1- i \omega)\Gamma(1- i \kappa \omega)} \exp\left( i b \omega - i\kappa \omega \log ( - i \omega)\right)\,,
\end{split}
\end{equation}
with 
\begin{equation}
    b= \kappa ( 1- \log(\kappa)) -\log(2)\,.
\end{equation}
\section{TBA Techniques } \label{sec:tba}
Polyakov and Wiegmann \cite{polyakov1983theory, polyakov1984goldstone, wiegmann1985exact}  showed in the 80s that it is possible to compute the free energy of an integrable system with a chemical potential $h$ turned on using a thermodynamic Bethe ansatz (TBA) technique.   Using these techniques, Hasenfratz, Niedermayer and Maggiore \cite{hasenfratz1990exact1, hasenfratz1990exact2} showed in 1990\footnote{This computation was intially performed for the $O(N)$ model, but was later also completed for Gross-Neveu models \cite{Forgacs:1991rs, Forgacs:1991ru} and PCM models \cite{Balog:1992cm, Hollowood:1994np}.} that it is possible to calculate the mass gap in integrable models by comparing the result from TBA with conventional Lagrangian pertubation theory.  Building from this we will will apply, in section  \ref{sec:PertTBA},  the techniques pioneered by \cite{volin2010mass, volin2011quantum, Marino:2019eym, Marino:2019wra} to extract an  expansion for the free energy of $\lambda$-model in $\frac{1}{h}$     the large order behaviour of which we will study extensively in section \ref{sec:Asymp}.

\subsection{Free Energy}
To present the TBA equations we will specialise to the case described above in which we introduce a chemical potential $h$ such that only a single particle dominates the ensemble at large $h$.\footnote{That we can reduce the TBA system to involve just one species of particle from the fundamental representation singled out by the applied chemical potential is of course an assumption that makes the problem readily tractable.  One anticipates that states of higher mass and higher charge are energetically disfavoured, but properly speaking this assumption ought to be proven starting from a complete nested TBA system (which we do not attempt here).} With  $K(\theta)$  the appropriate  scattering kernel, the  TBA equations determine the density distribution of states, $\chi(\theta)$,  via
\begin{equation} \label{eq:tbaintegralequation}
 m \cosh(\theta)  = \chi(\theta) - \int_{-B}^B K(\theta-\theta')\chi(\theta')d\theta'\,, \qquad \theta^2<B^2\,, 
\end{equation}
from which the charge and energy density follow   
\begin{equation} \label{eq:tbaintegralequation2} 
    e =m\int_B^{-B}\chi(\theta)\cosh(\theta)\frac{d\theta}{2\pi}\,,  \quad 
    \rho =\int_B^{-B}\chi(\theta) \frac{d\theta}{2\pi}\,.
\end{equation}
A critical complexity of this system is that the occupied states lie within a Fermi surface specified by $B$, which is however a function of $h$ (with large $B$ corresponding to large $h$).   Supposing that we have calculated the energy density, thought of as a function of the charge density $e= e(\rho)$, then 
we can reconstruct a free energy density, $\mathcal{F}(h)$, from a Legendre transform: 
\begin{equation} \label{eq:LegendreFreeEnergy}
    \rho = - \mathcal{F}'(h) \, , \quad 
    \mathcal{F}(h) -\mathcal{F}(0) = e(\rho) - \rho h\,.
\end{equation}

\subsection{Resolvent Approach}
It will prove useful to recast the integral equation that determines $\chi(\theta)$ in terms of a resolvent function defined by   
\begin{equation} \label{eq:resolventdefn}
    R(\theta) = \int_{-B}^B \frac{\chi(\theta')}{\theta - \theta'} d\theta'.
\end{equation}
The resolvent is analytical everywhere except around the interval $[-B,B]$ where it has an ambiguity given by 
\begin{equation} \label{eq:resolvantambiguity}
    \chi(\theta) = -\frac{1}{2\pi i} \left( R^+(\theta ) - R^-(\theta  ) \right)\,,
\end{equation}
where we use the short hand notation $R^\pm(\theta ) = R(\theta  \pm i \epsilon) $.  Suppose that the kernel can be cast in terms of some  operator $\mathsf{O}$ as     $K(\theta) = \frac{1}{2\pi i }  \mathsf{O} \frac{1}{\theta}$, then the eq. \eqref{eq:tbaintegralequation} is equivalent to a Riemann-Hilbert problem
\begin{equation} \label{eq:RTBAshiftop}
    R^+(\theta) - R^-(\theta)  + \mathsf{O}  R(\theta) = -2\pi i m \cosh \theta\,.
\end{equation}
A determination of $R(\theta)$ is then equivalent to solving the TBA system and once known the charge density is immediately extracted as 
\begin{equation} \label{eq:rhoinR}
    \rho = - \frac{1}{2\pi} \text{Res}_{\theta = \infty}R(\theta)\,. 
\end{equation}
We briefly now review the approach of \cite{volin2010mass, volin2011quantum, Marino:2019eym} which does so by considering ansatz solutions for the resolvent in two limits (the {\em edge} and {\em bulk}) and matching them to fix all undetermined coefficients.

\subsubsection{Edge Ansatz}
We begin first with the edge limit in which the weak coupling limit  $B \rightarrow \infty$ is taken whilst keeping an edge coordinate $z = 2 (\theta - B)$ fixed and small.  This evidently scales to large $\theta$  and hence probes the properties of $\chi(\theta)$ around the vicinity of the Fermi energy, $B$.  This limit is best studied by considering the Laplace transform of the resolvent \eqref{eq:resolventdefn} given by
\begin{equation} \label{eq:Rlaps}
    R(z)= \int_0^\infty \widehat{R}(s) e^{-s z }ds \,,\qquad \quad  \widehat{R}(s) = \frac{1}{2 \pi i} \int_{-i \infty+ \delta }^{i \infty+ \delta } e^{s z } R(z) d z \,.
\end{equation}
Note at large $B$ the energy density is related to this Laplace transformation by  
\begin{equation} \label{eq:rhoinR2}
   e = \frac{m e^B}{4 \pi} \widehat{R}(1/2)\, . 
\end{equation}

The key result of \cite{Marino:2019eym, Marino:2019wra} is that in the edge limit the Laplace transformed resolvent has the following form
\begin{equation} \label{eq:edgeansatz}
    \widehat{R}(s) = m e^B \Phi(s) \frac{\Phi\left(\frac{1}{2}\right)}{2}\left(\frac{1}{s + 1/2} + Q(s)\right)\,, \quad \Phi(s) = G_+(2 i s)\,,
\end{equation}
where $G_+(s)$ is the WH decomposition \eqref{eq:WHDecomp} of the (Fourier transformed) scattering kernel and $Q(s)$ is a series in large $s$ and a perturbative expansion in $\frac{1}{B}$ of the form
\begin{equation} \label{eq:Qser}
    Q(s) = \frac{1}{Bs} \sum_{m,n=0}^\infty \frac{Q_{n,m}}{B^{m+n} s^n}\,.
\end{equation} 
It should be noted that the coefficients $Q_{n,m}$ may still depend on $\log B$. 

\subsubsection{Bulk Ansatz}
In the bulk limit we let $B\rightarrow \infty$ and $\theta \rightarrow \infty$ but we keep $u=\theta/B$ fixed, we are hence studying the regime where $\theta$ is in the bulk, between $0$ and $B$. The precise form of the Bulk ansatz depends on the model. For the $\lambda$-model, we shall take the same bulk ansatz for the Gross-Neveu model \cite{Marino:2019eym}, which is given by 
\begin{equation} \label{eq:bosonicbulk2}
    R(u) = \sum_{n=1}^\infty\sum_{m=0}^\infty\sum_{k=0}^{n+m} c_{n,m,k} \frac{u^{e(k+1)}}{B^{m+n} (u^2-1)^n} \left[\log \frac{u-1}{1+u}\right]^k\,,
\end{equation}
where $e(k)$ is $0$ if $k$ is even and $1$ if $k$ is odd.The bulk ansatz can be motivated by constructing it using functions that are analytic outside the interval $[-B,B]$, where they have a logarithmic branch cut.\footnote{This is different from the PCM bulk ansatz which also has a square root branch cut along the interval $[-B,B]$.} This is precisely the analytic structure demanded by eqs. \eqref{eq:resolventdefn} and \eqref{eq:resolvantambiguity}. 


\subsection{Matching}
 If we re-expand the bulk ansatz \eqref{eq:bosonicbulk2} in an edge regime where $z=2(\theta-B)$ is fixed, we should recover the expansion in the edge regime given by \eqref{eq:edgeansatz}. Here a miraculous feature occurs:  upon comparing expansions order by order in large $B$, then order by order in large $z$ (which is small $s$) and then in $\log(z)$, we can solve for all the coefficients $c_{n,m,k}$ and $Q_{n,m}$. One peculiarity of the procedure is that we perform this matching only for the regular terms of the expansion $z^{-n}$ ($n\geq 0$), while we disregard all divergent terms $z^{n}$ ($n>0$). Using a desktop PC, over the course of a week, we solved the system up to 38 orders.  
Once this calculation is completed, we compute $e$ and $\rho$. Using equations \eqref{eq:edgeansatz} and \eqref{eq:rhoinR} we can express $\rho$ and $e$ in terms of the coefficients by
\begin{equation}
\begin{split}    
    e &= \frac{m^2 e^{2B} \Phi(1/2)^2}{8\pi} \left[ 1 + \sum_{m=1}^\infty \frac{1}{B^m} \sum_{n=0}^{m-1} 2^{n+1} Q_{n,m-1-n} \right]\,, \\
    \rho &= 2 \pi \sum_{m=0}^\infty \frac{c_{1,m,0}}{B^m}\,.
    \end{split}
\end{equation} 
Explicitly the first few coefficients required to determine up to order $B^{-2}$ are given by 
\begin{align}\nonumber
       c_{1,0,0}=4 \sqrt{\kappa }  \, , \quad  c_{1,1,0}= -2 \kappa ^{3/2}  \, , \\ 
\nonumber        c_{1,2,0} =  \frac{1}{2} \kappa
   ^{3/2} ( 2 - \kappa - 4 \log 2 + 4 \kappa \log(2 B/\kappa )) \, ,\\ 
       Q_{0,0} =0 \, , \quad Q_{1,0}=0 \,, \quad   Q_{0,1} = \frac{\kappa}{4} \, . 
    \end{align}
 The last step is to calculate the quantity $\frac{e}{\rho^2}$ as an expansion in $B$ the first terms of which are 
\begin{equation} \label{eq:eoverrhoinB}
 \frac{8 \kappa}{\pi} \frac{e}{\rho^2} = 1+ \frac{\kappa}{B} + \frac{\kappa}{B^2}\left(1 - \log(2) + \frac{\kappa}{2} + \kappa \log(2 B/\kappa) \right) + {\cal O}(B^{-3})\, .
\end{equation}
As this result depends on $\log(B)$, it is convenient to define a new effective coupling $\gamma$ in terms of which the perturbative expansion is free from logarithms as we shall do in the next section.

\subsection{Perturbative result} \label{sec:PertTBA}

Before introducing the log-free coupling, we show our results are consistent with those of \cite{appadu2017quantum}, which determines the mass gap of this theory. Using standard TBA techniques, they find an expansion for the free energy given by 
\begin{equation} \label{eq:HollowoodF}
\begin{split}
    \mathcal{F}(h) -\mathcal{F}(0) &= - \frac{2 h^2 \kappa}{\pi} \Big\{ 1 - 2\kappa \alpha
    + 2 \kappa \alpha^2 \big[2+ \kappa + \log 4 + + 2\kappa \log\kappa + 2 \kappa \log \alpha \big] \\
&- 8 \kappa^2 \alpha^3 \log(\alpha)\big[  ( -2+2 \kappa + \log 4 + 2 \kappa \log(\kappa) +\kappa \log(\alpha) \big]  + \mathcal{O}(\alpha^3)\Big\}\,.
\end{split}
\end{equation}
The coupling $\alpha$ is here defined by 
\begin{equation} \label{eq:hollowoodalpha}
    \frac{1}{\alpha} = 2 \log \left(\frac{2 h }{m} \sqrt{\frac{8 \kappa }{\pi}}\right)\,.
\end{equation}
By using the Legendre transformation \eqref{eq:LegendreFreeEnergy} we can compute the total energy $e$ from eq. \eqref{eq:HollowoodF}.  
Doing so, we obtain the expression
\begin{equation} \label{eq:Hollowoodevarrho}
\begin{split}
   \frac{8 \kappa}{\pi} \frac{e}{\rho^2} = &1+2 \alpha  \kappa - 2 \kappa\alpha ^2  (2 \kappa \log
   (\alpha  \kappa)- \kappa -2+\log (4)) +\\&8 \kappa^2 \alpha ^3
   \big[\kappa \log^2 (\alpha ) + (\log(\alpha)-1)(-2 \log (4) + 2 \kappa \log (\kappa)) \big] + \mathcal{O}\left(\alpha ^4\right)\, .
\end{split}
\end{equation}
From eq. \eqref{eq:eoverrhoinB}, it follows that $\frac{e}{\rho^2} = \chi_0 + \mathcal{O}(\alpha)$ where $\chi_0 = \frac{\pi}{8 \kappa}$. Therefore to leading order we have $h = \frac{\partial e}{\partial \rho} = 2 \chi_0 \rho$, which leads to $\rho = \frac{4 h \kappa}{\pi}$. Looking at eq. \eqref{eq:hollowoodalpha}, we should thus define a coupling by
\begin{equation}
    \frac{1}{\alpha} = 2 \log \left(\frac{\rho}{m} \sqrt{\frac{2 \pi}{\kappa}}\right)\,.
\end{equation}
This defines $\alpha$ in terms of $B$. Inverting the relation and substituting into the series \eqref{eq:eoverrhoinB} recovers precisely the expansion \eqref{eq:Hollowoodevarrho}, providing an important consistency check for our programme.

We now take inspiration from the Gross-Neveu treatment of \cite{Marino:2019eym} to create a series expansions for $\frac{e}{\rho^2}$ that is log-free. This is appropriate because we have that to leading order $\Delta F \sim- h^2 + \mathcal{O}(\alpha)$, which leads to a coupling defined by\footnote{This is in contrast to the PCM calculation where the free energy has a structure $\Delta F  \sim - \frac{h^2}{\alpha} + \mathcal{O}(\alpha^0)$, which leads to a coupling $\frac{1}{\gamma} + (\xi-1) \log\gamma \propto \log \rho$.}
\begin{equation} \label{eq:lambdacoupling}
    \frac{1}{\gamma} + \xi \log \gamma = \log\frac{2 \pi \rho}{m /c}\,, \quad \xi = \frac{\beta_2}{\beta_1^2} = - \frac{k}{N} = - \kappa\,.
\end{equation}
One could demand that the right hand side be $\log \frac{2 \pi \rho}{\Lambda_{MS}}$, where $\Lambda_{MS}$ is the cut-off in the minimal subtraction scheme. To achieve this one has to tune the constant $c= c_{MS}$ such that   $c_{MS}\Lambda_{MS} = m$.  A key outcome of \cite{appadu2017quantum} determines that $c_{MS} = e^{3/2} N^{-1/2}$.   However, we shall exercise the freedom 
to pick a $c$ of our own choosing,
\begin{equation} \label{eq:lambdapseudomassgap}
    c= \frac{2^{-\kappa}\Gamma(\kappa)}{\pi}\,,
\end{equation}
such that resulting expressions appear considerably simplified.   This leads to an expansion that is log-free in the coupling, given by
\begin{equation} \label{eq:lambdapertser}
\begin{split}
    \frac{8 \kappa}{\pi} \frac{e}{\rho^2}  &= \sum_{n=0}^\infty a_n \gamma^n =  1 + \kappa\gamma + \frac{\kappa}{2}[2- \kappa] \gamma^2 + \\ 
    &  \frac{\kappa}{2}\left[ 3  -5\kappa   + 2\kappa^2 \right] \gamma^3 + \frac{\kappa}{8} \left[ 3(8-\zeta (3)) - 61 \kappa + 52\kappa^2-15\kappa^3\right] \gamma^4 + \\
   & \frac{\kappa}{12}\left[90 -18
   \zeta (3) +\kappa  (33 \zeta (3)-288)+355 \kappa ^2-203 \kappa ^3 + 46 \kappa ^4 \right]\gamma^5 + \\
   & \frac{\kappa}{32} \big[45  (16- 4 \zeta (3)- \zeta (5))+2 \kappa (259 \zeta (3)-1338)+\frac{1}{3}
   \kappa ^2 (12274-1329 \zeta (3)) \\
   &-3285 \kappa ^3+1412 \kappa ^4-\frac{787 \kappa
   ^5}{3} \big] \gamma^6+ \mathcal{O}(\gamma^7)\, .
\end{split}
\end{equation}
In the next Section we shall explore this perturbative expansion further.
\section{Asymptotic Analysis} \label{sec:Asymp}

In this Section, we will quantitatively analyse the 38 orders of the perturbative series obtained in the previous Section. The goal shall be to compute an asymptotic formula for the growth of the coefficients as a function of $\kappa$. After obtaining such a formula, we can compute its Borel ambiguity, which can later be compared against an ambiguity of a transseries. 

\begin{figure}[tp]
    \centering
        \subfigure{
        \includegraphics[height=1.5in]{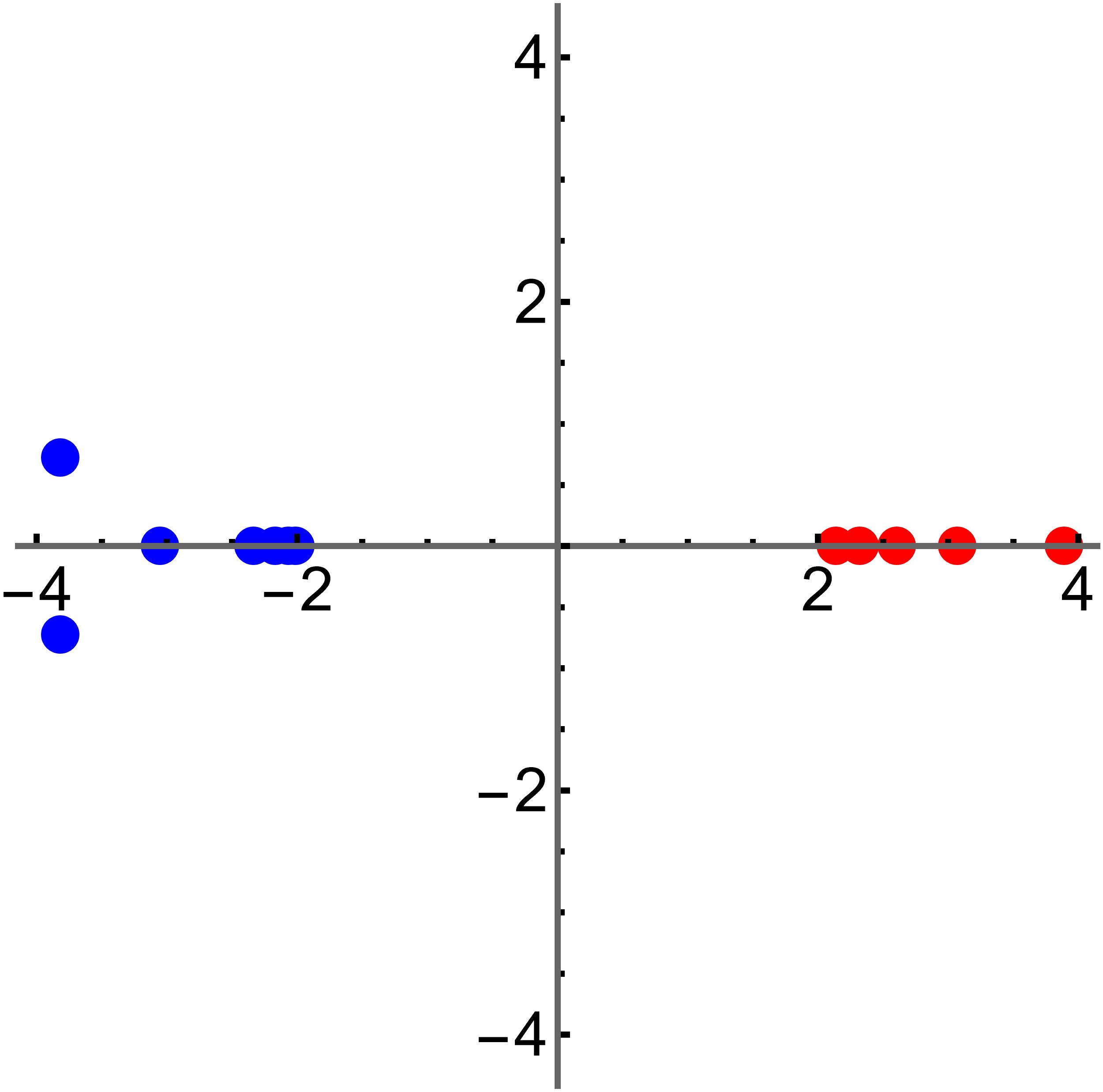}
        }
        \subfigure{
        \includegraphics[height=1.5in]{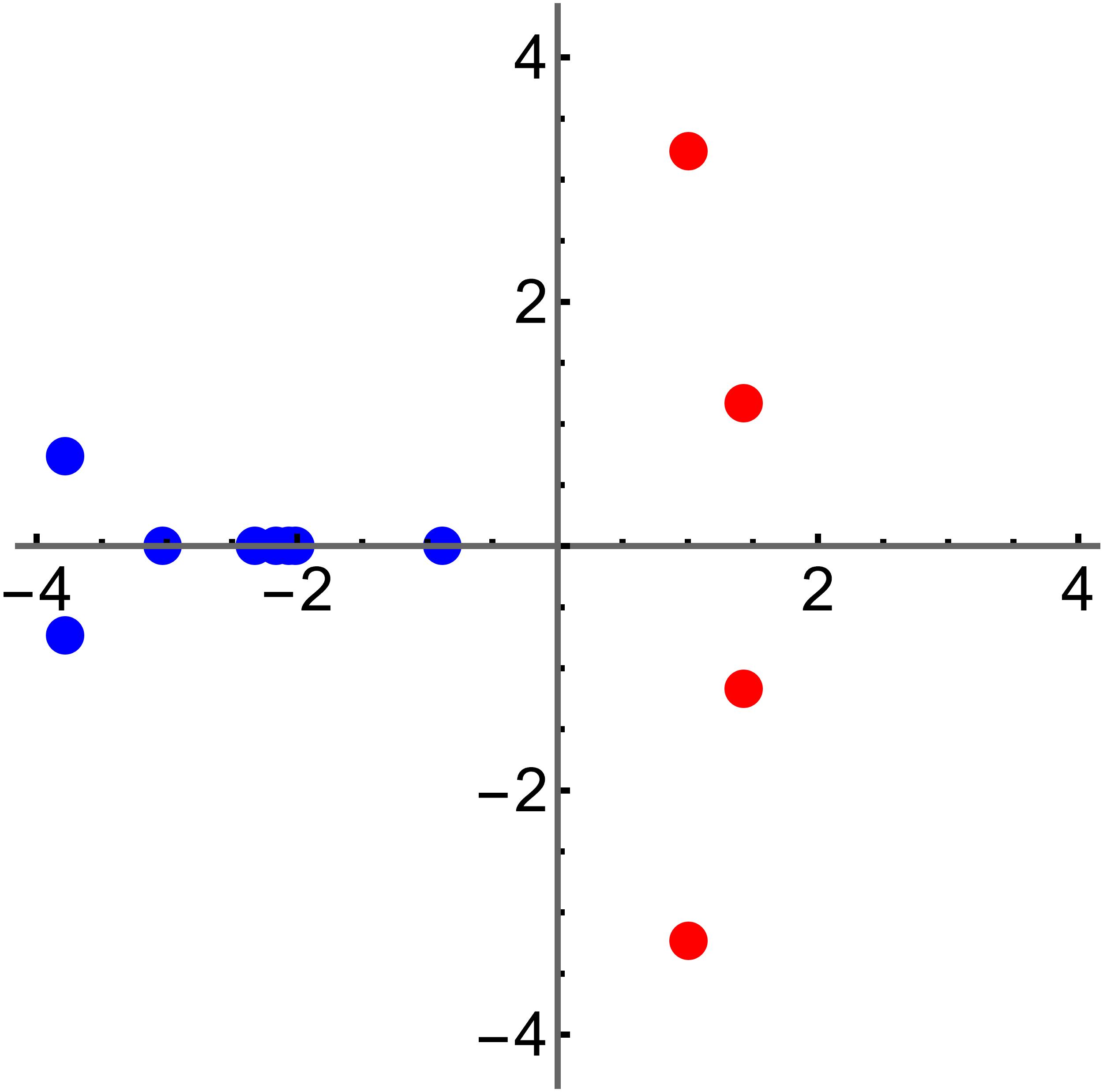}
        }  
        \subfigure{
        \includegraphics[height=1.5in]{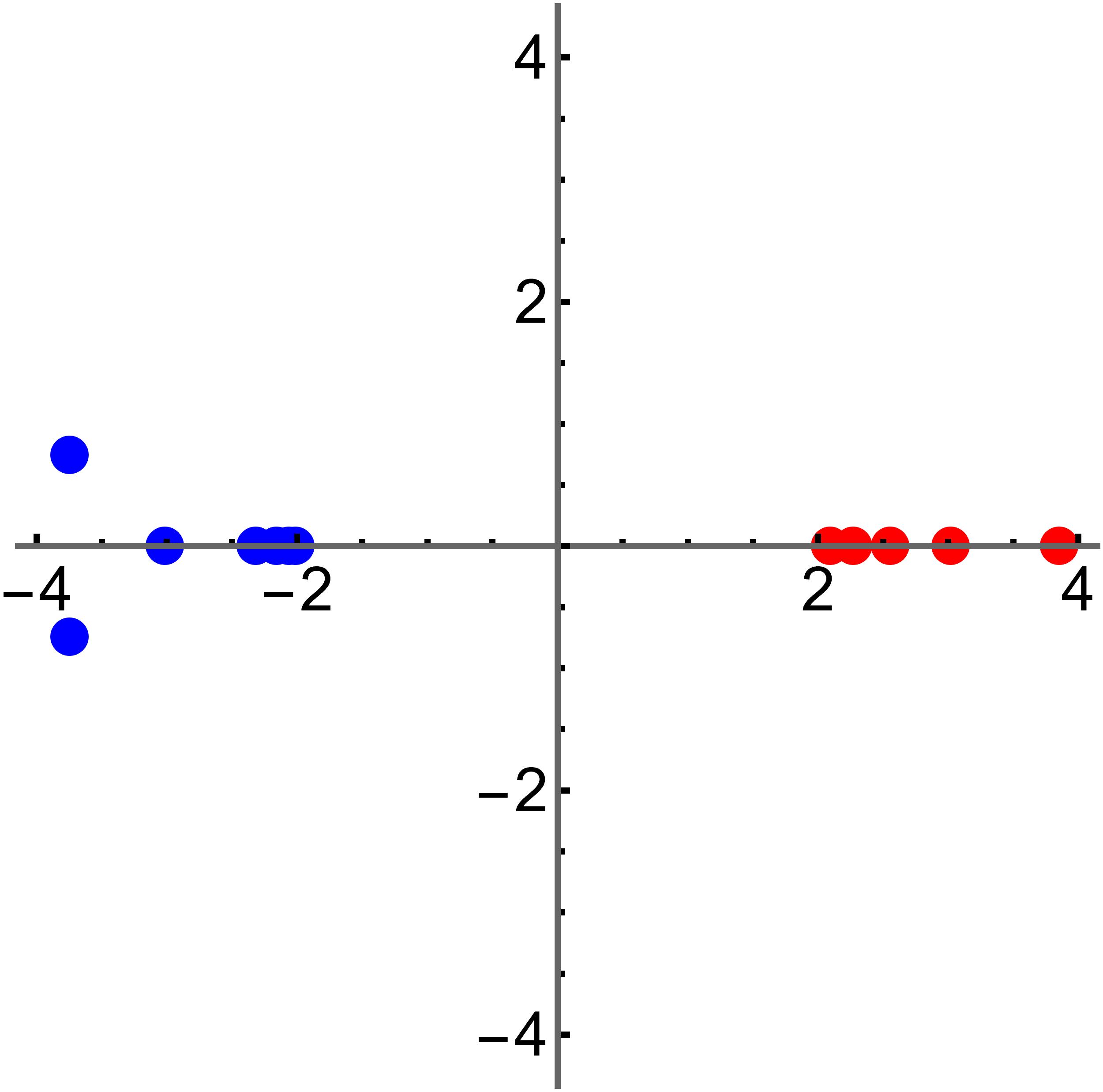}
        }  
    \caption{Left to right, for $\kappa=0.98$, $1$ and $1.02$,   the Borel-Pad\'{e}-poles in the $\zeta$-plane. Evident are singularities at $\zeta = \pm 2 $, with the positive pole removed for $\kappa =1$.}
    \label{fig:BorelPoles}
\end{figure}

As the perturbative series can readily be seen to exhibit factorial growth, as a first step to resummation we introduce the Borel transform 
\begin{equation}  \label{eq:Boreltransform}
{\cal B}\left[   \frac{8 \kappa}{\pi} \frac{e}{\rho}^2 \right] \equiv  \sum_{n=0}^\infty \frac{a_n}{n!}\zeta^n\, . 
\end{equation}
This series has a finite radius of convergence but typically has either, or both, poles and branch cuts.   The pole/branch point closest to the origin in the $\zeta$ plane is governed by the leading asymptotic behaviour.  Of course, numerically one does not have all orders with which to establish this Borel transformation, rather only a finite number of coefficients $a_n$ for $n<N$ say.   Here the Borel-Pad\'{e} method can be employed: we compute 
$ {\cal B}_N [ \frac{8 \kappa}{\pi} \frac{e}{\rho}^2 ] = \sum_{n=0}^N \frac{a_n}{n!}\zeta^n  = \frac{P (\zeta)}{Q(\zeta)} + \mathcal{O}(\zeta)^{N+1}$
in which $P$ and $Q$ are polynomials in $\zeta$ of degree $N/2$.  This results in a picture in which an accumulation of poles (i.e. zeros of $Q$) is indicative of a branch point.   We perform this numerically for various values of $\kappa$ and generically we find evidence of branch points at $\zeta = \pm 2$  whose location is independent of $\kappa$ except that for $\kappa \in \mathbb{Z}_{>0}$ the pole in the positive axis is removed - see Figure \ref{fig:BorelPoles}.  Pole/ branch points in the negative real axis of the Borel plane indicate contributions to $a_n$ of alternating sign whereas the contributions to $a_n$ that result in poles on the  positive axis would have the same sign.  Here the analysis indicates that we have both.  With 38 perturbative coefficients this analysis should only be regarded as indicative but is sufficient to inform an educated guess as to the asymptotic behaviour of the $a_n$ which we will robustly verify below. 

\begin{figure}[tp]
    \centering
     \subfigure{
        \includegraphics[height=1.6in]{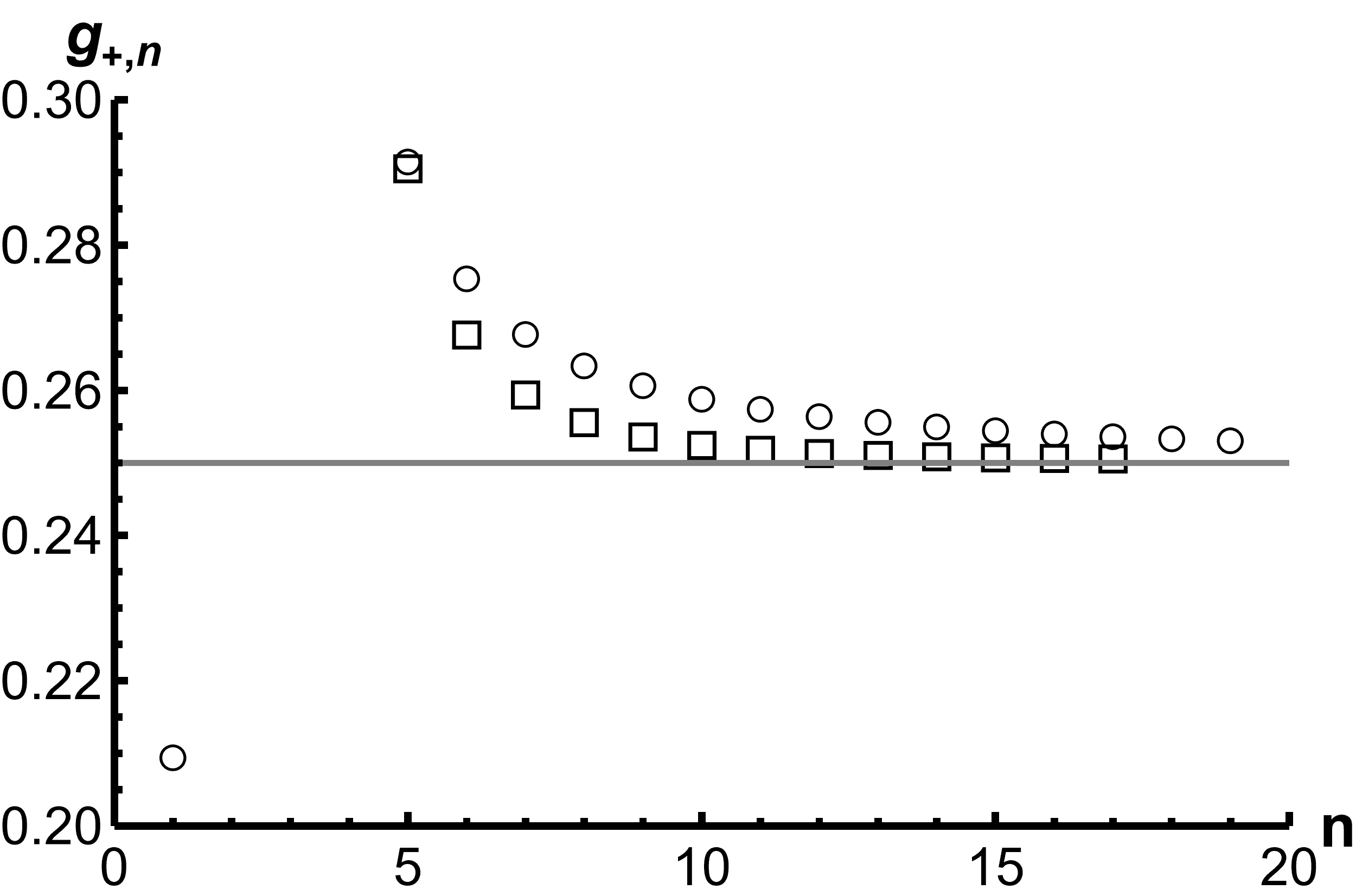}
        }
     \subfigure{
        \includegraphics[height=1.6in]{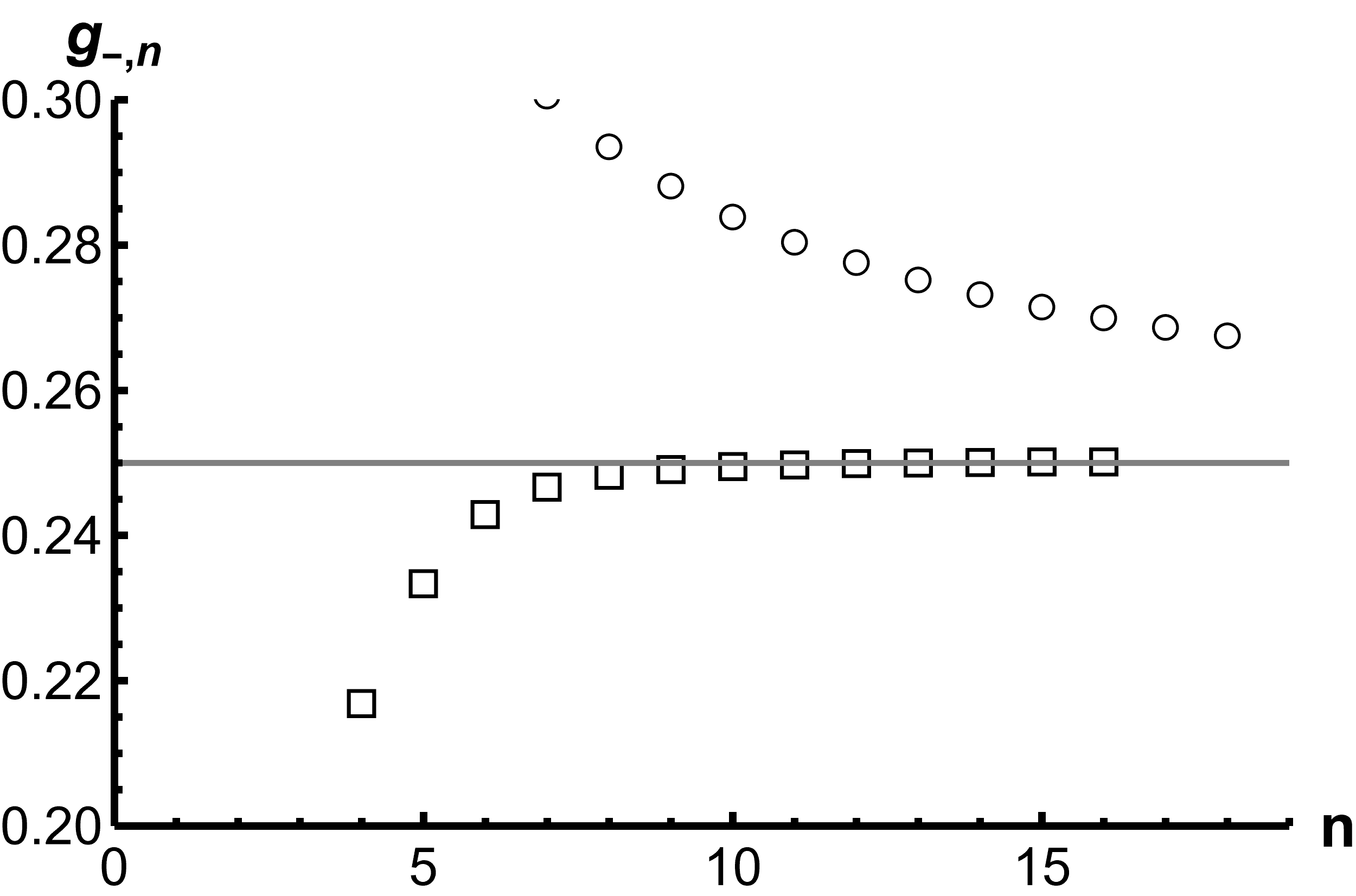}
        }
    \caption{The series  $g_{+,n}$ (left) and $g_{-,n}$ (right) given by eq. \eqref{eq:gpmseries} displayed for $\kappa=0.6$.  Circle markers indicate the raw data, square markers the second Richardson transformation with accelerated convergence.  The final values of the second Richardson transform differ by $0.11\%$ and $0.05\%$ respectively from the expected value $\frac{1}{4}$.}
    \label{fig:BorelpoleRT}
\end{figure}

Motivated by the Borel-Pad\'{e} analysis we assume the coefficients grow, to leading approximation, as
\begin{equation}
    a_n \approx A_+\Gamma(n+1)/S^n + A_- \Gamma(n+1)/(-S)^n + \mathcal{O}(n^{-1}) \,. 
\end{equation}
A first verification is to establish the factor $S$ which can be done noting that 
\begin{equation} \label{eq:gpmseries}
    g_{+,n}:=\frac{ a_{2n }}{ 2n(2n-1) a_{2n-2  }} \approx \frac{1}{S^2}\,, \qquad g_{-,n}:=\frac{ a_{2n+1}}{ 2n(2n-1) a_{2n-1}} \approx \frac{1}{S^2}\,.
\end{equation}
 We find, see Figure   \ref{fig:BorelpoleRT},  that the series $g_{\pm,n}$ converge to $\frac{1}{4}$, independent of $\kappa$ thus establishing $S = 2$ in accordance with the expectation from the Borel-Pad\'{e} analysis. 

Having established the factorially growing character of the perturbative series, we now propose a more refined ansatz for the $a_n$.   Our central claim can be summarised by stating that the perturbative series has coefficients that have a leading large order behaviour as
\begin{equation} \label{eq:gammasubleading}
    a_n \approx \frac{A_+}{2^n}\sum_{l=0}^\infty \beta^+_l \Gamma(n + a_+ -l) + \frac{A_-}{(-2)^n} \sum_{l=0}^\infty \beta^-_l \Gamma(n + a_- -l)\,,
\end{equation} 
where we normalise $\beta_0^\pm = 1$ and the first few coefficients are 
\begin{equation} \label{eq:asympresults}
\begin{split}
    a_\pm &= \mp 2 \kappa \, , \\
    A_\pm &= \frac{8^{\pm 1}}{\pi} \sin( \mp \kappa)  \frac{\Gamma(\pm\kappa)}{\Gamma(\mp  \kappa)}= - \frac{8^{\pm1}}{\Gamma(\mp \kappa) \Gamma(1 \mp \kappa)}\, ,\\ 
    \beta^-_1 &= -\beta^-_2 = -4 \kappa \, . 
\end{split}
\end{equation}

\begin{figure}[tp]
    \centering
        \subfigure{
        \includegraphics[height=1.5in]{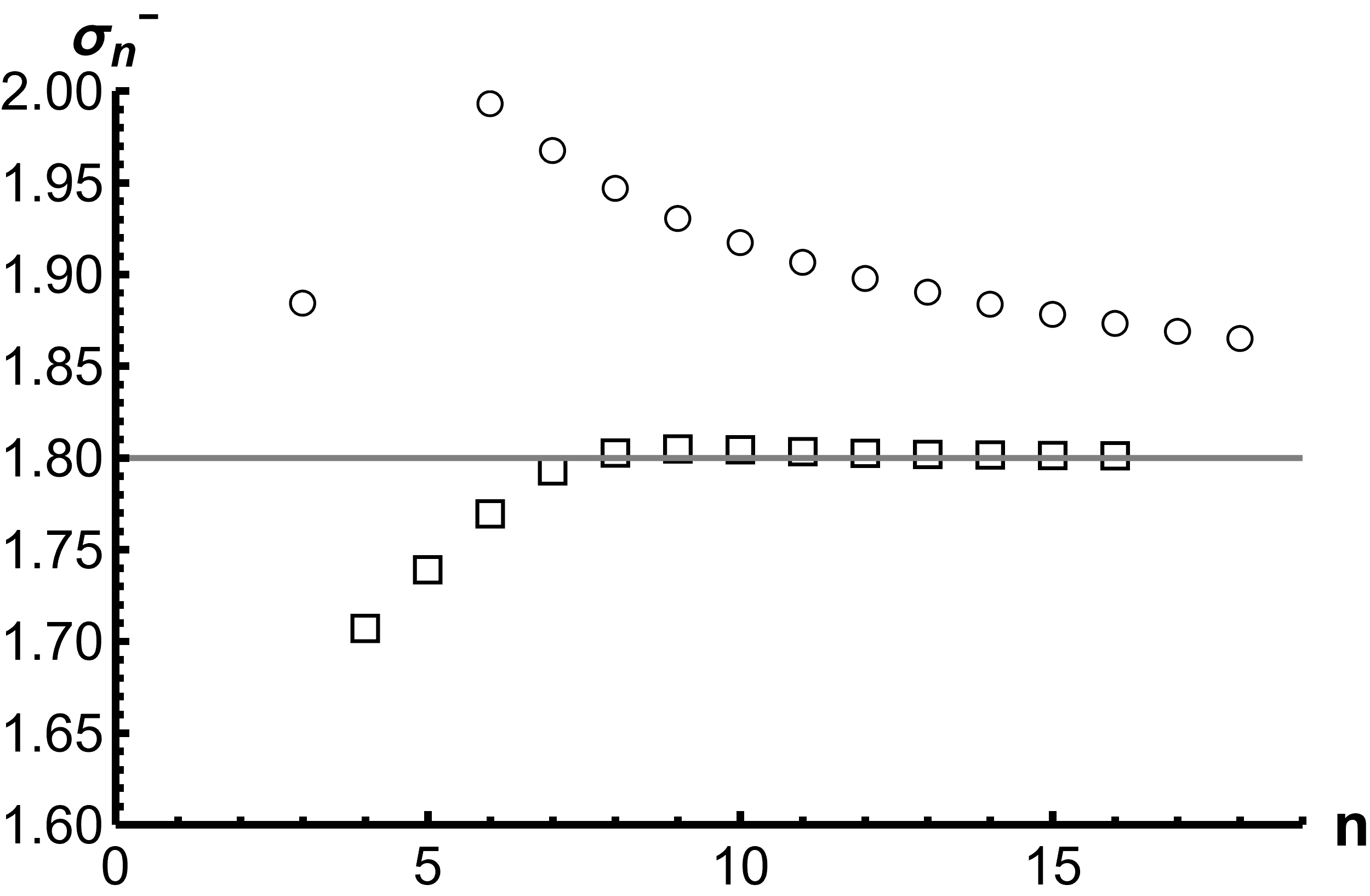}
        }
        \subfigure{
        \includegraphics[height=1.5in]{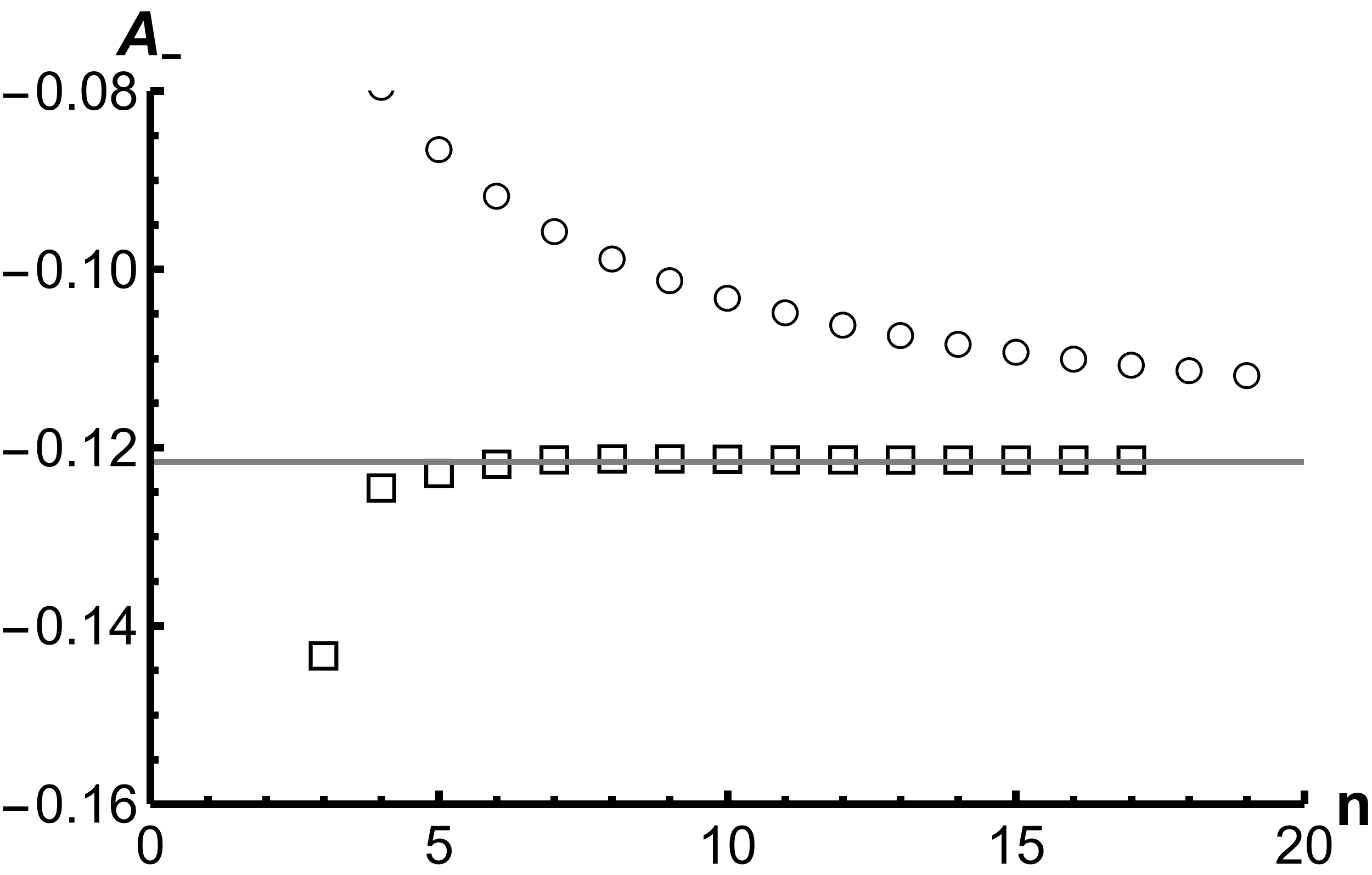}
        }  
    \caption{The series $\sigma_n^-$ (left) converges to $a_-$ using \eqref{eq:defnauxsigman}.  Using eq. \eqref{eq:fnasymps} we display (right) the sequence that converges to $A_-$.   Circle markers indicate the raw data, square markers the second Richardson transformation.  For both, we display results for $\kappa=0.9$. The second Richardson transform converge to the expected results given by eq \eqref{eq:asympresults} up to  errors of  $0.011\%$ and $0.00068\%$ respectively.}
    \label{fig:RTamAm}
\end{figure}

\begin{figure}[tp]
    \centering
        \subfigure{
        \includegraphics[height=1.6in]{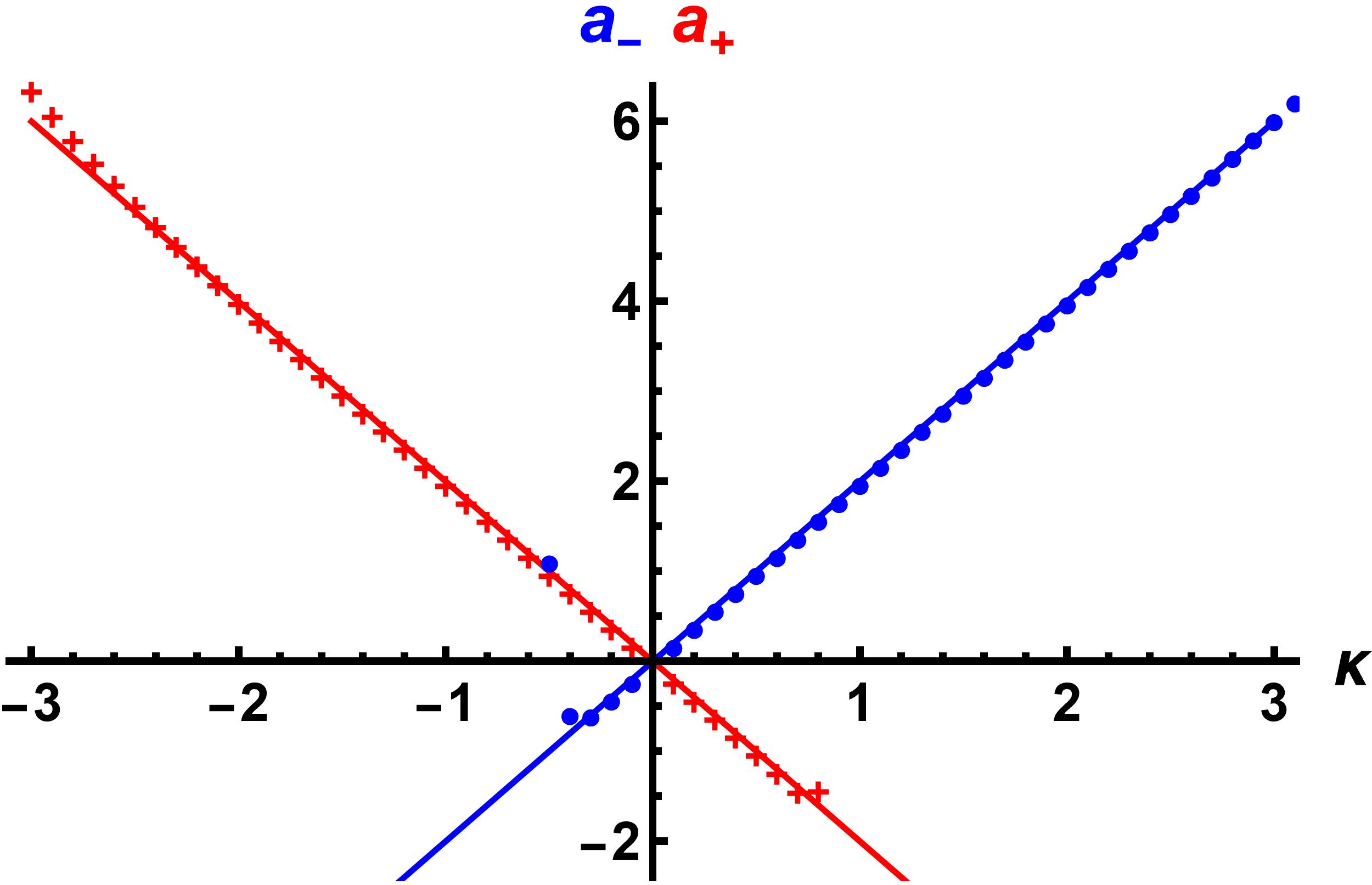}
        }  
        \subfigure{
        \includegraphics[height=1.6in]{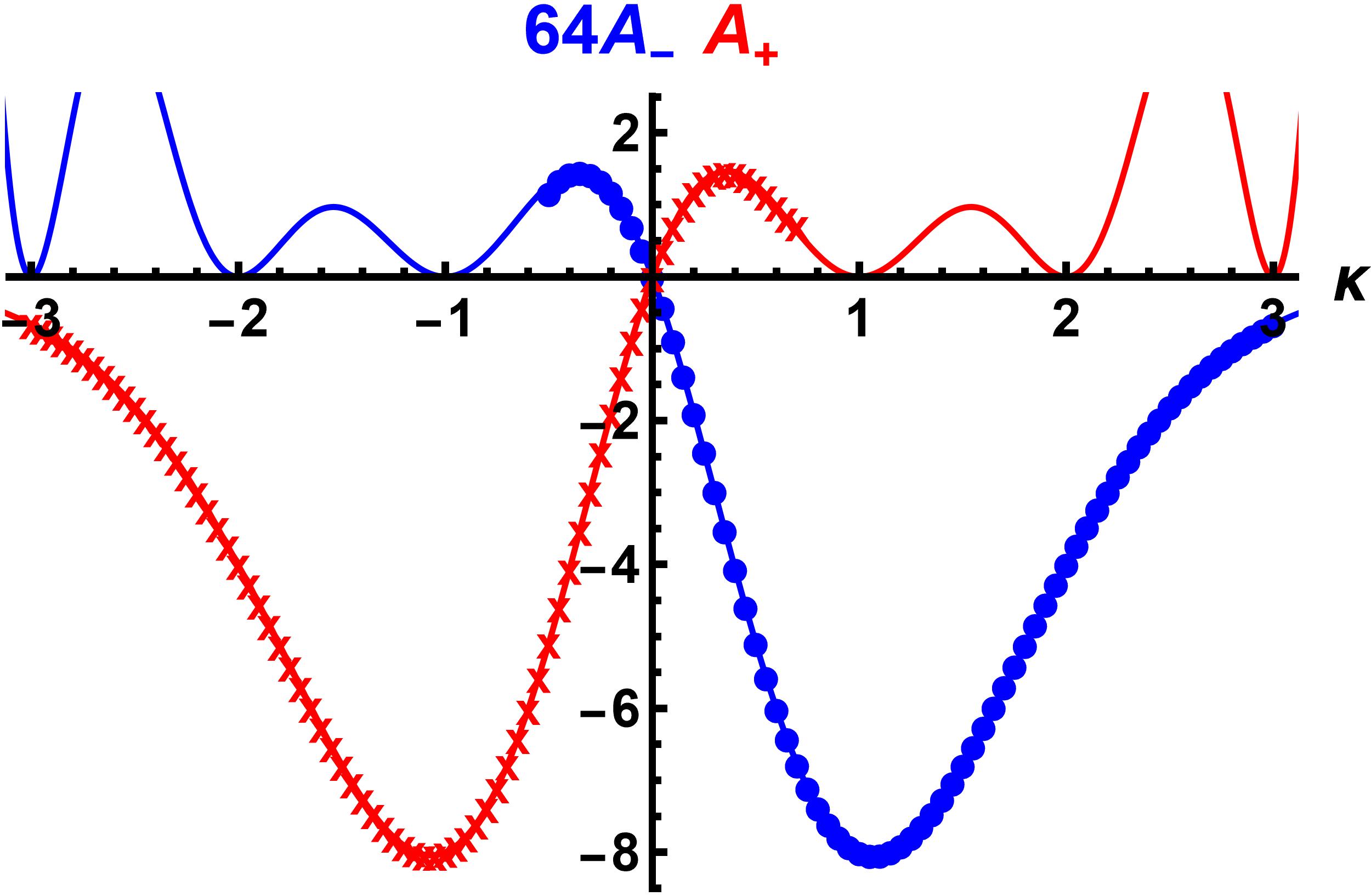}
        }
    \caption{The second Richardson transformation of the sequences \eqref{eq:defnauxsigman} (left) and \eqref{eq:fnasymps} (right) to determine $a_\pm$ and $A_\pm$ as functions of $\kappa$. $a_+, A_+$ are indicated by  red crosses  and $a_-, A_+$  by blue points with solid lines showing the analytic formula of  eq. \eqref{eq:asympresults}. }
    \label{fig:analyticapm}
\end{figure}

To support these claims, we shall define the auxiliary series
\begin{equation} \label{eq:defnauxcn}
    c_n = \frac{2^n}{\Gamma(n+1)} a_n\,,
\end{equation}
to take care of the leading factorial and geometric growth. 
We project to the alternating and non-alternating parts of the series by considering
\begin{equation} \label{eq:defnauxfk}
    f^\pm_k = c_{2k} \pm c_{2k-1}\,,
\end{equation}
which have asymptotics
\begin{equation} \label{eq:fnasymps}
    f^\pm_n = 2 A_\pm (2n)^{a_\pm-1}\Big( 1+ \mathcal{O}\Big(\frac{1}{n}\Big)\Big)\,,
\end{equation}
such that the sequences
\begin{equation} \label{eq:defnauxsigman}
   \sigma^\pm_n = 1 + n \log \frac{f^\pm_{n+1}}{f^\pm_n}\,,
\end{equation}
converge to $a_\pm $. With $a_\pm$ determined one can then directly consider the asymptotics of $f^\pm$ to establish $A_\pm$.   Figure \ref{fig:RTamAm} illustrates the convergence of this procedure for a fixed value of $\kappa$, and  Figure \ref{fig:analyticapm} establishes the functional form of these coefficients for various values of $\kappa$.

A methodological subtlety is that, from empirical observation, the contributions from the alternating sector, i.e. $A_-$ (and associated  subleading terms), are dominant for $\kappa >0 $ over those of the non-alternating $A_+$ sector.   Thus to extract the non-alternating contributions we first establish the leading alternating contribution as described above and then repeat the process working instead with a new series in which the leading alternating contribution has been subtracted.  However, when $\kappa$ becomes sufficiently large, the sub-leading  alternating contribution becomes comparable to that of the leading non-alternating contribution.   This limits the reliability of determination numerically of the $A_+, a_+$ coefficients to small values of $\kappa$.  However, these coefficients can be more readily verified by continuing to the $\kappa<0$ regime where they are more dominant. 

\begin{figure}[tp]
    \centering
    \subfigure{\includegraphics[height=1.4in]{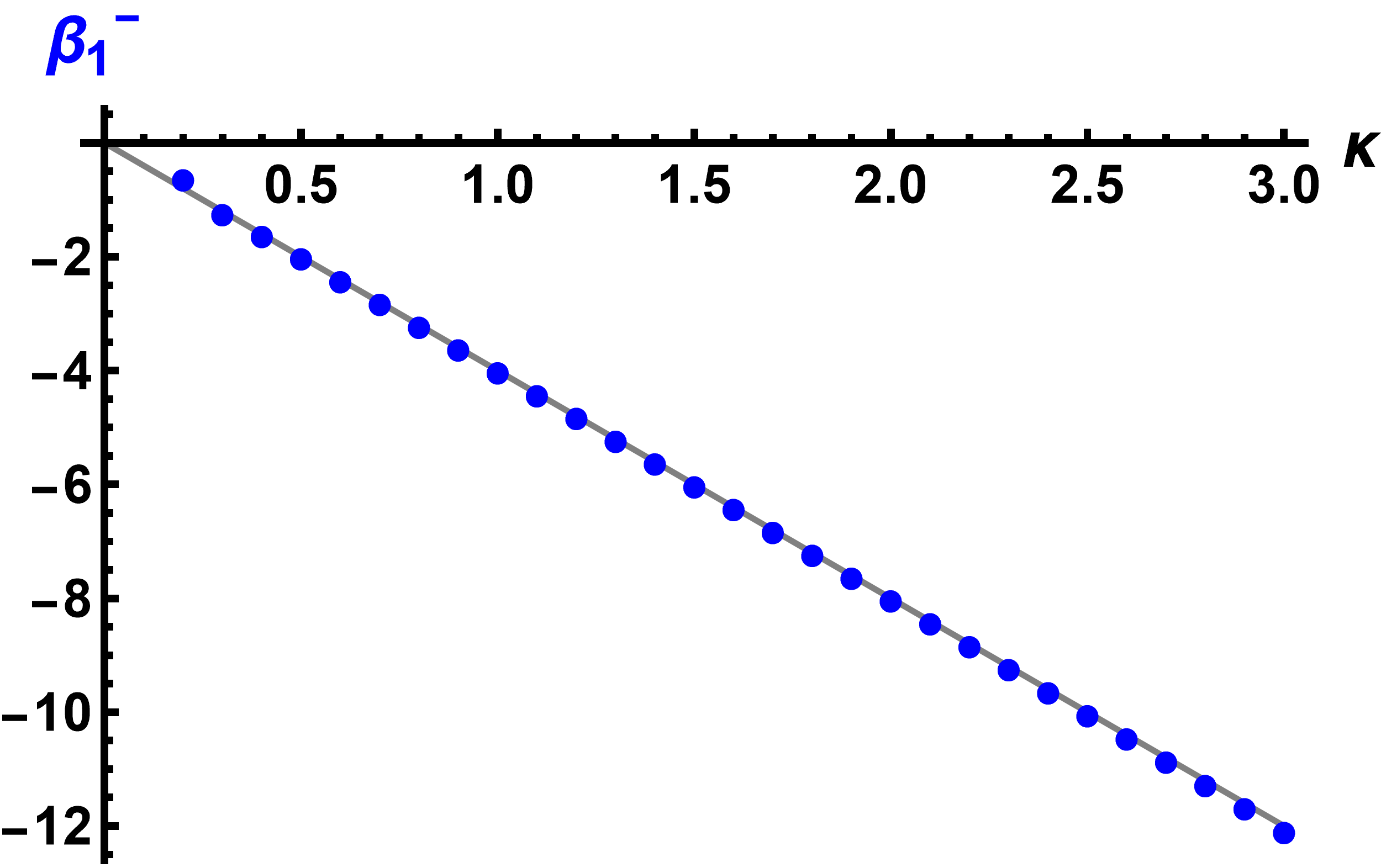}}
    \subfigure{\includegraphics[height=1.4in]{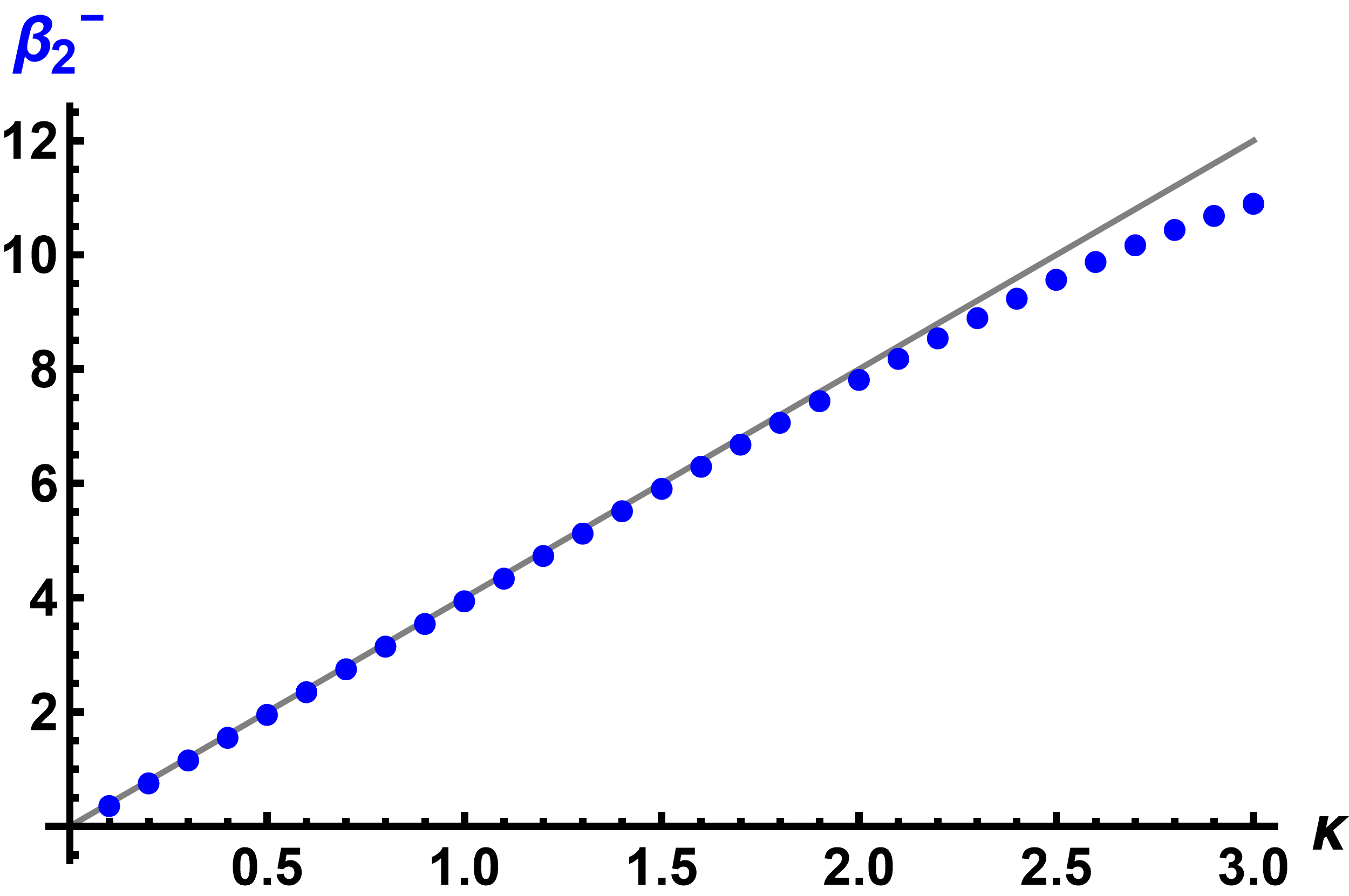}}
    \caption{The sub-leading coefficient $\beta_1^-$ (left) and  $\beta_2^-$ (right) for various values of $\kappa$.  Shown is the terminal value of the second Richardson Transformation of the sequence that gives $\beta_n^-$ constructed from $f_n$ after subtraction of leading alternating and non-alternating asymptotics.  Grey lines correspond to   $ \beta_1^- = - 4 \kappa$ and $ \beta_2^- = + 4 \kappa$.  A noticeable drift in $\beta_2^-$ for larger values of $\kappa$ suggests pollution from further sub-dominant terms contributing at this order of perturbation theory.}
    \label{fig:betafunctional}
\end{figure}

Having determined in this fashion the leading contributions to $a_n$, these can then be subtracted from the data, the analysis repeated {\em mutatis mutandis}, to determine the sub-leading $\beta_k$ coefficients (and again for similar reasons to the above the $\beta^-_k$ coefficients are more readily extracted). Figure  \ref{fig:betafunctional} gives the numerical form of $\beta_1^-$ and $\beta_2^-$ as a function of $\kappa$ indicating a linear relationship. 

\begin{figure}[tp]
    \centering
    \subfigure{\includegraphics[height=1.6in]{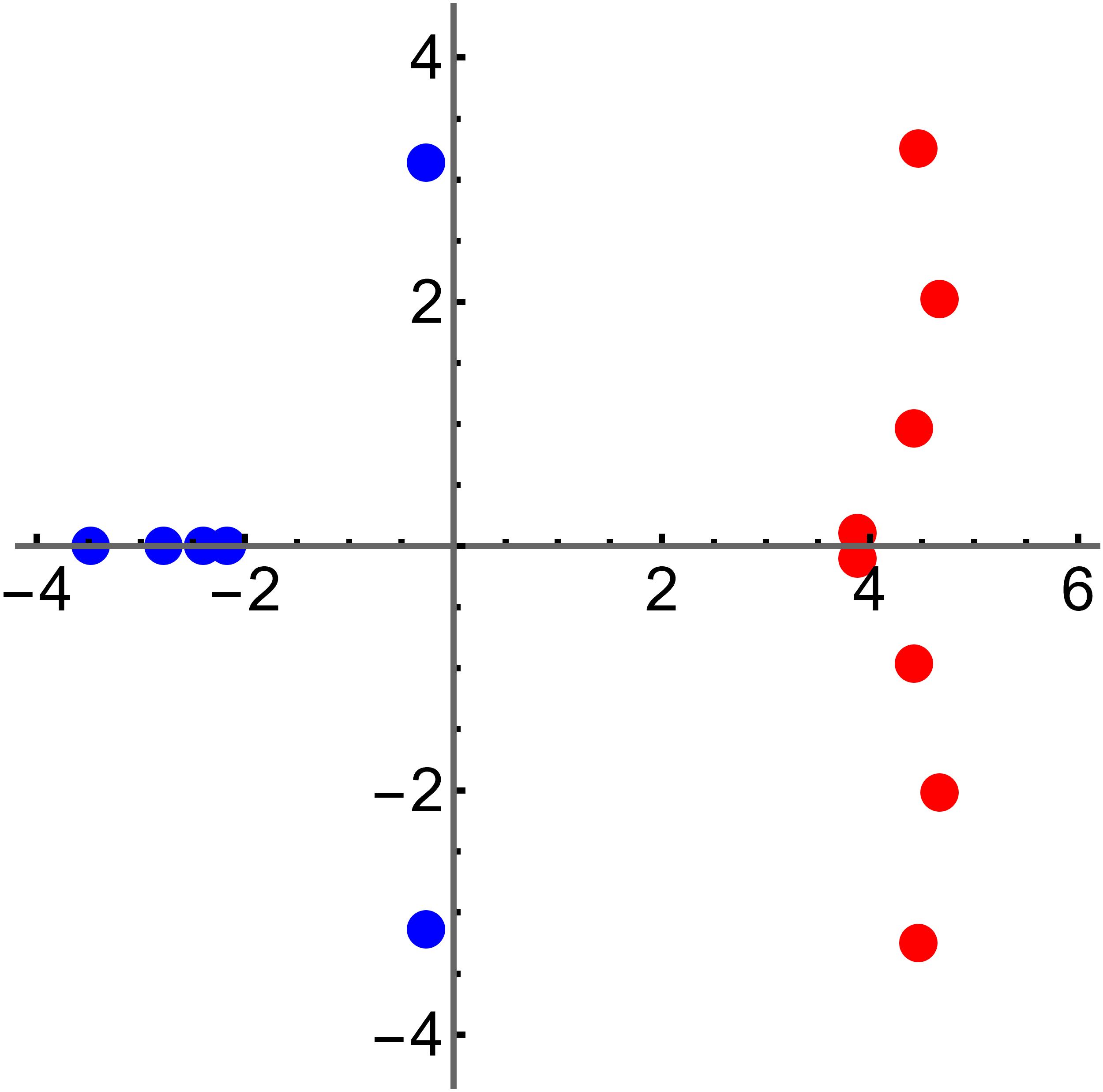}}
    \subfigure{\includegraphics[height=1.6in]{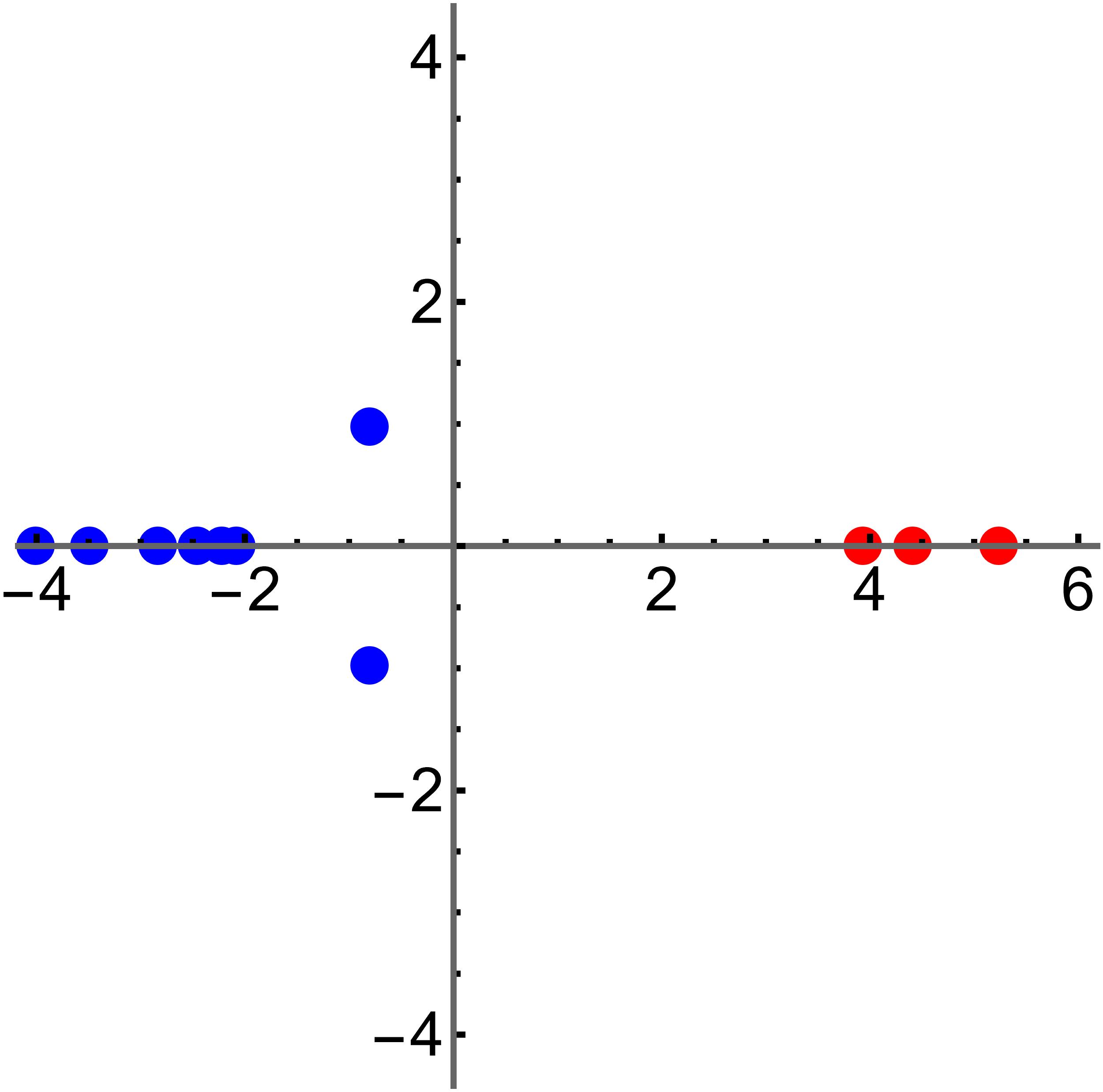}}
    \caption{After subtracting the leading alternating and non-alternating contributions, we again perform a Borel-Pad\'{e} computation for $\kappa = - 0.75$ (left) and $\kappa = 0.4$ (right). This seems to suggest that there is no longer a Borel singularity at $\zeta=2$, but instead finding one at $\zeta=4$.}
    \label{fig:subleadingBorel}
\end{figure}

It becomes somewhat challenging to extract further subleading contributions from the data available.  However, one can consider defining a new series, $\tilde{a}_n$, comprised by taking the data set and subtracting the already established asymptotic form of eq.~ \eqref{eq:asympresults}.   Using the Borel-Pad\'{e} again to this subtracted series produces some evidence, see Figure \ref{fig:subleadingBorel}, of a compelling feature.   Instead of poles at $\zeta = \pm 2$, as would be anticipated should the ansatz \eqref{eq:gammasubleading}, one finds that leading positive pole appears to be at $\zeta = +4$.   The interpretation here is that the subtraction has removed the entire non-alternating terms with behaviour $2^{-n}$, suggesting that all fluctuations $\beta_{n>0}^+ = 0$ and the next contribution comes with twice the ``action'' $4^{-n}$.  

This behaviour is in accordance  with the Parisi-'t Hooft conjecture  \cite{Parisi:1978az, Parisi:1978bj, tHooft:1977xjm};  the leading poles in the Borel plane at  $\zeta = \pm 2 $   lie at integer values and  the values of $a_\pm = \mp 2 \kappa = \pm 2\xi$ are as expected (see \cite{Marino:2019eym}).\footnote{We thank M Mari\~{n}o and T Reis for illuminating us on this point.}   The pole at $\zeta=+2$ is accordingly  interpreted as an IR renormalon.   A similar procedure of subtraction (removing the IR renormalon) used above (in Figure \ref{fig:subleadingBorel}) was performed in \cite{Marino:2021dzn} to expose new Borel renormalon poles that were {\em not} in accordance with Parisi-'t Hooft  in cases including e.g. the Gross-Neveu model.   Here however, Figure \ref{fig:subleadingBorel} indicates that the next most proximate IR renormalon pole is found in a location that {\em are} consistent with Parisi-'t Hooft.

\section{Transseries and Ambiguity Cancellation} \label{sec:transseries}

In this section we compute the leading ambiguity of $\frac{e}{\rho^2}$ in two different ways. First, we calculate the Borel ambiguity of the large order behaviour of the perturbative sector established in \ref{sec:Asymp}. This is compared against an approach which solves the TBA system in terms of a transseries.

\subsection{Borel resummation and Large Order Perturbative Ambiguity} 
Naively, one could try to resum the original asymptotic series by performing a Laplace transform on the  Borel transform \eqref{eq:Boreltransform}  
\begin{equation}
   \frac{1}{\gamma} \int_0^\infty \mathcal{B}\left[ \frac{8 \kappa}{\pi} \frac{e}{\rho^2}\right] e^{-\zeta /\gamma} d \zeta = \frac{1}{\gamma} \int_0^\infty \sum_{n=0}^\infty \frac{a_n}{n!} \zeta^n e^{-\zeta/\gamma} \simeq  \sum_{n=0}^\infty a_n \gamma^n = \frac{8 \kappa}{\pi} \frac{e}{\rho^2}\,.
\end{equation} 
However, as we have seen, the Borel transform $\mathcal{B}\left[ \frac{8 \kappa}{\pi} \frac{e}{\rho^2}\right]$ generically has singularities along the positive real axis obstructing the contour of this integral. Therefore, we shall introduce a directional Borel resummation given by
\begin{equation}
   S_\theta\left[ \frac{8 \kappa}{\pi} \frac{e}{\rho^2}\right] = \frac{1}{\gamma} \int_0^{e^{i \theta} \infty} \mathcal{B}\left[ \frac{8 \kappa}{\pi} \frac{e}{\rho^2}\right] e^{-\zeta /\gamma} d \zeta \,.
\end{equation}
This procedure results, when integrating along a line without singularities, in a finite answer, which however, depends on the sign of $\theta$, i.e. there is an ambiguity in the resummation of the perturbative series.   This ambiguity, which is a Stokes phenomenon,  is studied by considering $S_{+\epsilon} - S_{-\epsilon}$.   This can be done analytically by using, instead of the numerically obtained results, a series whose coefficients are exactly the asymptotic form $a_n$ given by eq. \eqref{eq:gammasubleading} for all values of $n$:
\begin{equation} \label{eq:IRambiguity}
\begin{split}
    (S_{+\epsilon} - S_{-\epsilon})\left[ \frac{8 \kappa}{\pi} \frac{e}{\rho^2}\right] (\gamma) &= 2 \pi i A_+ \Big(\frac{2}{\gamma}\Big)^{a_+} e^{-2/\gamma} \sum_{k=0}^\infty \beta_k^+ \Big(\frac{\gamma}{2}\Big)^k  \\
    &= - \frac{16 \pi i}{\Gamma(-\kappa) \Gamma(1-\kappa)} \left(\frac{\gamma}{2}\right)^{2 \kappa} e^{-2/\gamma}[ 1+ \mathcal{O}(\gamma)]\,.
\end{split}
\end{equation}
Similarly, across the negative real axis we find a leading ambiguity given by
\begin{equation} \label{eq:UVambiguity}
\begin{split}
    (S_{\pi+\epsilon} - S_{\pi-\epsilon}) \left[ \frac{8 \kappa}{\pi} \frac{e}{\rho^2}\right](\gamma) &= 2 \pi i A_- \Big(-\frac{2}{\gamma}\Big)^{a_-} e^{2/\gamma} \sum_{k=0}^\infty \beta_k^- \Big(-\frac{z}{2}\Big)^k \\
    &= - \frac{ \pi i }{4 \Gamma(\kappa) \Gamma(1+\kappa)} \left(-\frac{\gamma}{2}\right)^{-2 \kappa} e^{2/\gamma}[ 1+ \mathcal{O}(\gamma)]\,.
\end{split}
\end{equation}

In these expressions we note the presence of an exponentially small parameter,   $ \sqrt{q_\gamma} = \left( \frac{2}{\gamma}\right)^{2 \kappa} e^{-2/\kappa}$  (the square root is for convenience later) characteristic of non-perturbative physics.   The main thrust of the modern resurgence paradigm is that physical quantities, here $e / \rho^2$, should be understood as a transseries, i.e. an   expansion in $  \sqrt{q_\gamma}$ whose terms are each formal (asymptotic) series in $\gamma$.  It is critical that whilst resummation may be ambiguous when applied to any individual term in this (here the perturbative $\sqrt{q_\gamma}^0$ sector), taken altogether the final result is non-ambiguous.  In particular, and this goes back to the pioneering work of Bogomol'nyi and Zinn-Justin  \cite{bogomolny1980calculation, zinn2004multi1, zinn2004multi2, ZinnJustin:1980uk}, the ambiguity of this perturbative sector should be compensated by a leading order ambiguity in an appropriate non-perturbative sector.  In the next section we shall verify that such an ambiguity cancellation does take place. 

\subsection{Transseries and Leading Non-Perturbative Ambiguity} 

In this series we shall apply a different type of analysis to the TBA equations which results in a transseries solution. 
The starting point shall be a reformulation of the TBA system as an integral equation for an auxiliary function $u(\omega)$,
\begin{equation} \label{eq:baseinteqnu}
     u(\omega) = \frac{i}{\omega} +\frac{1}{2\pi i}\int_{-\infty}^\infty  d\omega'   \frac{ e^{2i B \omega'} \varrho(\omega') u(\omega') }{\omega' + \omega+ i \delta }  \,,
\end{equation} 
where
\begin{equation} \label{eq:varrhodefn2}
\varrho(\omega) = \frac{ 1-i\omega} {   1+i\omega} \frac{G_-(\omega) }{G_+(\omega) }  \, , 
\end{equation}
together with the boundary condition 
\begin{equation} \label{eq:u(i)bc}
    u(i) = \frac{m}{2 h} e^{B} \frac{G_+(i) } {G_+(0)}\,.
\end{equation}
Having established the function $u$, the free energy is given by
\begin{equation} \label{eq:freeenergyinu} 
        \Delta F(h) =   -\frac{1}{2\pi}  h^2 u(i) G_-(0)^2   \left( 1 -   \frac{1}{2\pi i} \int_{-\infty}^\infty d\omega \frac{e^{2 i\omega B } u(\omega) \varrho(\omega)}{\omega -i}    \right)  \,.  
\end{equation}

We will now apply to the $\lambda$-model the techniques pioneered by \cite{Marino:2021dzn} to solve this recursively order by order in a perturbative parameter and a non-perturbative parameter. 
The idea is to move the integration contour of the integral equation \eqref{eq:baseinteqnu} into the UHP so that it envelops all the branch cuts and poles in the UHP. The Sine-Gordon model is special as it only has poles but no branch cut.  This was studied in \cite{Zamolodchikov:1995xk} and gives rise to a convergent rather than asymptotic expansion. However, in the case of the $\lambda$-deformed model, we are dealing with both poles and a branch cut along the imagine axis of $\rho(\omega)$. 

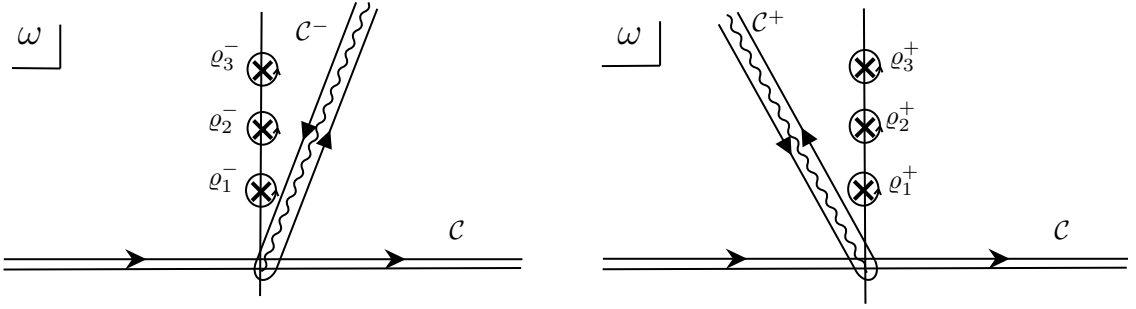
\begin{figure}[tp]
    \centering

\tikzset{every picture/.style={line width=0.75pt}} 

\begin{tikzpicture}[x=0.75pt,y=0.75pt,yscale=-1,xscale=1]

\draw    (152.62,23.78) -- (151.85,166.68) ;
\draw    (282.18,152.58) -- (23.81,153.34) ;
\draw  [line width=1.5]  (147.85,109.62) -- (156.93,118.69)(156.93,109.62) -- (147.85,118.69) ;
\draw    (52.12,52.23) -- (52.12,30.15) ;
\draw    (153.38,148.16) -- (23.95,148.16) ;
\draw [shift={(95.17,148.16)}, rotate = 180] [fill={rgb, 255:red, 0; green, 0; blue, 0 }  ][line width=0.08]  [draw opacity=0] (10.72,-5.15) -- (0,0) -- (10.72,5.15) -- (7.12,0) -- cycle    ;
\draw    (282.8,148.16) -- (153.38,148.16) ;
\draw [shift={(224.59,148.16)}, rotate = 180] [fill={rgb, 255:red, 0; green, 0; blue, 0 }  ][line width=0.08]  [draw opacity=0] (10.72,-5.15) -- (0,0) -- (10.72,5.15) -- (7.12,0) -- cycle    ;
\draw    (199.81,18.95) -- (150.23,148.64) ;
\draw [shift={(173.23,88.46)}, rotate = 290.92] [fill={rgb, 255:red, 0; green, 0; blue, 0 }  ][line width=0.08]  [draw opacity=0] (8.93,-4.29) -- (0,0) -- (8.93,4.29) -- cycle    ;
\draw    (159.82,153.52) -- (210.37,22.19) ;
\draw [shift={(186.89,83.19)}, rotate = 111.05] [fill={rgb, 255:red, 0; green, 0; blue, 0 }  ][line width=0.08]  [draw opacity=0] (8.93,-4.29) -- (0,0) -- (8.93,4.29) -- cycle    ;
\draw  [draw opacity=0] (159.82,153.52) .. controls (157.9,157.45) and (154.2,159.54) .. (151.56,158.19) .. controls (148.91,156.84) and (148.32,152.57) .. (150.23,148.64) -- (155.02,151.08) -- cycle ; \draw   (159.82,153.52) .. controls (157.9,157.45) and (154.2,159.54) .. (151.56,158.19) .. controls (148.91,156.84) and (148.32,152.57) .. (150.23,148.64) ;  
\draw   (144.84,113.54) .. controls (144.84,109) and (148.19,105.32) .. (152.32,105.32) .. controls (156.45,105.32) and (159.79,109) .. (159.79,113.54) .. controls (159.79,118.08) and (156.45,121.76) .. (152.32,121.76) .. controls (148.19,121.76) and (144.84,118.08) .. (144.84,113.54) -- cycle ;
\draw    (157.7,116) -- (159.79,113.54) ;
\draw    (160.99,116.44) -- (159.79,113.54) ;

\draw  [line width=1.5]  (148.62,78.4) -- (157.69,87.47)(157.69,78.4) -- (148.62,87.47) ;
\draw   (145.6,82.33) .. controls (145.6,77.79) and (148.95,74.11) .. (153.08,74.11) .. controls (157.21,74.11) and (160.56,77.79) .. (160.56,82.33) .. controls (160.56,86.87) and (157.21,90.55) .. (153.08,90.55) .. controls (148.95,90.55) and (145.6,86.87) .. (145.6,82.33) -- cycle ;
\draw    (158.46,84.78) -- (160.56,82.33) ;
\draw    (161.75,85.22) -- (160.56,82.33) ;

\draw  [line width=1.5]  (148.62,48.71) -- (157.69,57.78)(157.69,48.71) -- (148.62,57.78) ;
\draw   (145.6,52.64) .. controls (145.6,48.1) and (148.95,44.42) .. (153.08,44.42) .. controls (157.21,44.42) and (160.56,48.1) .. (160.56,52.64) .. controls (160.56,57.18) and (157.21,60.86) .. (153.08,60.86) .. controls (148.95,60.86) and (145.6,57.18) .. (145.6,52.64) -- cycle ;
\draw    (158.46,55.09) -- (160.56,52.64) ;
\draw    (161.75,55.53) -- (160.56,52.64) ;

\draw    (205.91,18.99) .. controls (206.85,21.15) and (206.24,22.7) .. (204.08,23.64) .. controls (201.92,24.58) and (201.31,26.13) .. (202.24,28.29) .. controls (203.18,30.45) and (202.57,32) .. (200.41,32.95) .. controls (198.25,33.89) and (197.64,35.44) .. (198.58,37.6) .. controls (199.52,39.76) and (198.91,41.31) .. (196.75,42.25) .. controls (194.59,43.19) and (193.98,44.74) .. (194.92,46.9) .. controls (195.86,49.06) and (195.25,50.61) .. (193.09,51.55) .. controls (190.92,52.49) and (190.31,54.04) .. (191.25,56.21) .. controls (192.19,58.37) and (191.58,59.92) .. (189.42,60.86) .. controls (187.26,61.8) and (186.65,63.35) .. (187.59,65.51) .. controls (188.53,67.67) and (187.92,69.22) .. (185.76,70.16) .. controls (183.6,71.11) and (182.99,72.66) .. (183.93,74.82) .. controls (184.87,76.98) and (184.26,78.53) .. (182.1,79.47) .. controls (179.94,80.41) and (179.33,81.96) .. (180.26,84.12) .. controls (181.2,86.28) and (180.59,87.83) .. (178.43,88.77) .. controls (176.27,89.72) and (175.66,91.27) .. (176.6,93.43) .. controls (177.54,95.59) and (176.93,97.14) .. (174.77,98.08) .. controls (172.61,99.02) and (172,100.57) .. (172.94,102.73) .. controls (173.88,104.89) and (173.27,106.44) .. (171.11,107.38) .. controls (168.94,108.32) and (168.33,109.87) .. (169.27,112.04) .. controls (170.21,114.2) and (169.6,115.75) .. (167.44,116.69) .. controls (165.28,117.63) and (164.67,119.18) .. (165.61,121.34) .. controls (166.55,123.5) and (165.94,125.05) .. (163.78,125.99) .. controls (161.62,126.94) and (161.01,128.49) .. (161.95,130.65) .. controls (162.89,132.81) and (162.28,134.36) .. (160.12,135.3) .. controls (157.96,136.24) and (157.35,137.79) .. (158.28,139.95) .. controls (159.22,142.11) and (158.61,143.66) .. (156.45,144.6) .. controls (154.29,145.55) and (153.68,147.1) .. (154.62,149.26) .. controls (155.56,151.42) and (154.95,152.97) .. (152.79,153.91) -- (152.62,154.35) -- (152.62,154.35) ;
\draw    (28,52) -- (52.12,52.23) ;
\draw    (453.27,22) -- (454.04,170.56) ;
\draw    (584.47,152.42) -- (322.85,153.2) ;
\draw  [line width=1.5]  (448.45,108.92) -- (457.64,118.1)(457.64,108.92) -- (448.45,118.1) ;
\draw    (351.52,50.81) -- (351.52,28.46) ;
\draw    (322.43,51.02) -- (351.52,50.81) ;
\draw    (454.04,147.94) -- (322.99,147.94) ;
\draw [shift={(395.02,147.94)}, rotate = 180] [fill={rgb, 255:red, 0; green, 0; blue, 0 }  ][line width=0.08]  [draw opacity=0] (10.72,-5.15) -- (0,0) -- (10.72,5.15) -- (7.12,0) -- cycle    ;
\draw    (585.09,147.94) -- (454.04,147.94) ;
\draw [shift={(526.07,147.94)}, rotate = 180] [fill={rgb, 255:red, 0; green, 0; blue, 0 }  ][line width=0.08]  [draw opacity=0] (10.72,-5.15) -- (0,0) -- (10.72,5.15) -- (7.12,0) -- cycle    ;
\draw    (380.77,30.01) -- (448.78,153.05) ;
\draw [shift={(417.2,95.91)}, rotate = 241.07] [fill={rgb, 255:red, 0; green, 0; blue, 0 }  ][line width=0.08]  [draw opacity=0] (8.93,-4.29) -- (0,0) -- (8.93,4.29) -- cycle    ;
\draw    (458.82,148.82) -- (390.18,23.95) ;
\draw [shift={(422.09,82)}, rotate = 61.2] [fill={rgb, 255:red, 0; green, 0; blue, 0 }  ][line width=0.08]  [draw opacity=0] (8.93,-4.29) -- (0,0) -- (8.93,4.29) -- cycle    ;
\draw  [draw opacity=0] (458.82,148.82) .. controls (458.82,148.82) and (458.82,148.82) .. (458.82,148.82) .. controls (460.61,152.87) and (459.81,157.09) .. (457.04,158.26) .. controls (454.27,159.43) and (450.57,157.1) .. (448.78,153.05) -- (453.8,150.94) -- cycle ; \draw   (458.82,148.82) .. controls (458.82,148.82) and (458.82,148.82) .. (458.82,148.82) .. controls (460.61,152.87) and (459.81,157.09) .. (457.04,158.26) .. controls (454.27,159.43) and (450.57,157.1) .. (448.78,153.05) ;  
\draw   (445.4,112.89) .. controls (445.4,108.3) and (448.79,104.57) .. (452.97,104.57) .. controls (457.15,104.57) and (460.54,108.3) .. (460.54,112.89) .. controls (460.54,117.49) and (457.15,121.22) .. (452.97,121.22) .. controls (448.79,121.22) and (445.4,117.49) .. (445.4,112.89) -- cycle ;
\draw    (458.42,115.38) -- (460.54,112.89) ;
\draw    (461.75,115.83) -- (460.54,112.89) ;

\draw  [line width=1.5]  (449.22,77.31) -- (458.41,86.5)(458.41,77.31) -- (449.22,86.5) ;
\draw   (446.17,81.29) .. controls (446.17,76.69) and (449.56,72.96) .. (453.74,72.96) .. controls (457.92,72.96) and (461.31,76.69) .. (461.31,81.29) .. controls (461.31,85.89) and (457.92,89.61) .. (453.74,89.61) .. controls (449.56,89.61) and (446.17,85.89) .. (446.17,81.29) -- cycle ;
\draw    (459.19,83.78) -- (461.31,81.29) ;
\draw    (462.52,84.22) -- (461.31,81.29) ;

\draw  [line width=1.5]  (449.22,47.25) -- (458.41,56.43)(458.41,47.25) -- (449.22,56.43) ;
\draw   (446.17,51.22) .. controls (446.17,46.63) and (449.56,42.9) .. (453.74,42.9) .. controls (457.92,42.9) and (461.31,46.63) .. (461.31,51.22) .. controls (461.31,55.82) and (457.92,59.55) .. (453.74,59.55) .. controls (449.56,59.55) and (446.17,55.82) .. (446.17,51.22) -- cycle ;
\draw    (459.19,53.71) -- (461.31,51.22) ;
\draw    (462.52,54.16) -- (461.31,51.22) ;

\draw    (384.79,25.31) .. controls (387.04,25.98) and (387.83,27.45) .. (387.16,29.71) .. controls (386.49,31.97) and (387.28,33.44) .. (389.54,34.11) .. controls (391.8,34.78) and (392.59,36.25) .. (391.91,38.51) .. controls (391.24,40.77) and (392.03,42.24) .. (394.29,42.91) .. controls (396.55,43.58) and (397.34,45.05) .. (396.66,47.31) .. controls (395.99,49.57) and (396.78,51.04) .. (399.04,51.71) .. controls (401.3,52.38) and (402.09,53.85) .. (401.41,56.11) .. controls (400.74,58.37) and (401.53,59.84) .. (403.79,60.51) .. controls (406.05,61.18) and (406.84,62.65) .. (406.16,64.91) .. controls (405.49,67.17) and (406.28,68.64) .. (408.54,69.31) .. controls (410.8,69.98) and (411.59,71.45) .. (410.91,73.71) .. controls (410.24,75.97) and (411.03,77.44) .. (413.29,78.11) .. controls (415.55,78.78) and (416.34,80.25) .. (415.66,82.51) .. controls (414.99,84.77) and (415.78,86.24) .. (418.04,86.91) .. controls (420.3,87.58) and (421.09,89.05) .. (420.41,91.31) .. controls (419.74,93.57) and (420.53,95.04) .. (422.79,95.71) .. controls (425.05,96.38) and (425.84,97.85) .. (425.16,100.11) .. controls (424.49,102.37) and (425.28,103.84) .. (427.54,104.51) .. controls (429.8,105.18) and (430.59,106.65) .. (429.91,108.91) .. controls (429.24,111.17) and (430.03,112.64) .. (432.29,113.31) .. controls (434.55,113.98) and (435.34,115.45) .. (434.66,117.71) .. controls (433.99,119.97) and (434.78,121.44) .. (437.04,122.11) .. controls (439.3,122.78) and (440.09,124.25) .. (439.41,126.51) .. controls (438.74,128.77) and (439.53,130.24) .. (441.79,130.91) .. controls (444.05,131.58) and (444.84,133.05) .. (444.16,135.31) .. controls (443.49,137.57) and (444.28,139.04) .. (446.54,139.71) .. controls (448.8,140.38) and (449.59,141.85) .. (448.91,144.11) .. controls (448.24,146.37) and (449.03,147.84) .. (451.29,148.51) .. controls (453.55,149.18) and (454.34,150.65) .. (453.66,152.91) -- (454.76,154.94) -- (454.76,154.94) ;

\draw (29.47,30) node [anchor=north west][inner sep=0.75pt]  [font=\Large]  {$\omega $};
\draw (168.43,26.02) node [anchor=north west][inner sep=0.75pt]    {$\mathcal{C}^{-}$};
\draw (244.99,127.51) node [anchor=north west][inner sep=0.75pt]    {$\mathcal{C}$};
\draw (124.27,97.87) node [anchor=north west][inner sep=0.75pt]    {$\varrho _{1}^{-}$};
\draw (124.27,68.18) node [anchor=north west][inner sep=0.75pt]    {$\varrho _{2}^{-}$};
\draw (125.03,36.2) node [anchor=north west][inner sep=0.75pt]    {$\varrho _{3}^{-}$};
\draw (328.72,30) node [anchor=north west][inner sep=0.75pt]  [font=\Large]  {$\omega $};
\draw (396.22,21.34) node [anchor=north west][inner sep=0.75pt]    {$\mathcal{C}^{+}$};
\draw (546.91,127.18) node [anchor=north west][inner sep=0.75pt]    {$\mathcal{C}$};
\draw (463.77,99.51) node [anchor=north west][inner sep=0.75pt]    {$\varrho _{1}^{+}$};
\draw (462.23,67.13) node [anchor=north west][inner sep=0.75pt]    {$\varrho _{2}^{+}$};
\draw (464.54,37.07) node [anchor=north west][inner sep=0.75pt]    {$\varrho _{3}^{+}$};

\end{tikzpicture}
    \caption{The contour $\mathcal{C} = (- \infty, \infty)$ is deformed into either of two ways. The branch cut, represented by the curvy line is moved to either the ray $C_+$ or $C_-$. In those cases respectively, the contour is deformed into $\mathcal{C}^+$ or $\mathcal{C}^-$. In both cases we pick up residues $\varrho_n^\pm$, but their values differ by the branch cut discontinuity. }
    \label{fig:branchcutdeformedcontour}
\end{figure}

To separate it from the poles, we  slightly move the cut away from the imaginary axis to the ray $C_\pm = \{ \xi e^{i \theta} |   \theta = \frac{\pi}{2} \pm \delta\}$.  By deforming the integration contour we isolate the contributions coming from the discontinuity over the cut and the residues at the poles (see Figure \ref{fig:branchcutdeformedcontour}).  As explained in \cite{Marino:2021dzn}  the choice of moving the branch cut to $C_+$ or $C_-$ is arbitrary and and gives rise to a leading non-perturbative ambiguity. Letting $\varrho_{n,\pm}$ be the residues at $x=x_n$ with the cut moved to $C_\pm$, after this contour pulling eq. \eqref{eq:baseinteqnu} becomes
 \begin{equation} \label{eq:integraleqnu}
       u(i x ) = \frac{1}{x} +\frac{1}{2\pi i} \int_{0}^{\infty e^{\pm i \epsilon}} dx' \,\frac{ e^{-2 B x'}u(i x'  )  \delta \varrho(i x' )  }{x' + x }       +     \sum_n   \frac{ e^{-2 B x_n}u_n \varrho_{n,\pm}  }{x_n + x }\,,
 \end{equation}
where $u_n \equiv u(i x_n)$ and $\delta \varrho$ is the discontinuity over the cut\footnote{ For the discontinuity function, we use the convention $\delta\rho(\omega) = \rho(\omega( 1 - i \epsilon) - \rho (\omega(1+ i \epsilon))$.}.



From the WH-decomposition \eqref{eq:WHlambda}, we evaluate $\varrho(\omega)$ using \eqref{eq:varrhodefn2} as
\begin{equation}
  \varrho(\omega) =  -\frac{\omega +i}{\omega-i} \frac{  \Gamma \left(\frac{i
   \omega }{2}+1\right)^2
   \Gamma (1-i \omega ) \Gamma
   (1-i \kappa  \omega
   )}{\Gamma
   \left(1-\frac{i \omega
   }{2}\right)^2 \Gamma (i
   \omega +1) \Gamma (i \kappa
    \omega +1)} e^{-2 i b
   \omega } e^{i \kappa \omega ( \log ( i \omega) + \log ( - i \omega))}\, .
\end{equation}
For generic values of $\kappa$, this has poles on the positive real axis at $\omega = i x_n = i \mu n$ with $\mu=2$ with residues given by
\begin{equation}  \label{eq:residuesvarrhopos}
\varrho_{n, \pm} =\mathop{\mathrm{Res}}_{x= x_n \pm i \epsilon} \varrho( i x)=
-2 i e^{2 n(2b \pm i \pi \kappa - 2 \kappa \log(2n))}n\frac{2n+1}{2n-1} \frac{((2n)!)^2}{(n!)^4}\frac{\Gamma(1+2n \kappa)}{\Gamma(1-2n \kappa)}\, . 
\end{equation} 
However, when $\kappa $ is rational some of these poles are removed. Suppose we express $\kappa \equiv \frac{k}{N} = p/q$ as a reduced fraction with $p,q$ coprime integers (i.e. $q = N / \text{gcd}(N,k)$), then the set of poles are located at $ x \in 2 \mathbb{N}  \setminus q \mathbb{N}$ (rather than $ x \in 2 \mathbb{N}$). Hence, the residue $\varrho_{n, \pm}$ evaluates to zero if $2n \in  2 \mathbb{N}  \cap q \mathbb{N}  $, i.e. $2n$ is a multiple of $q$. In particular, when $k$ is an integer multiple of a half, i.e. $q=1$ or $q=2$,  all poles are removed entirely.  
If $\varrho_1 = 0$, then $\varrho_n = 0$ for all $n$; in what follows we shall consider only the case where $\varrho_1 \neq 0$ which is most relevant to our discussion. 

The discontinuity function is given by
\begin{equation} \label{eq:deltavarrho}
     \delta \varrho(i x) = 2 i \frac{x+1}{x-1} e^{2b x} e^{-2\kappa x \log x } \sin(\kappa \pi x) \frac{\Gamma(1-x/2)^2 \Gamma(1+ x) \Gamma(1 +\kappa x) } {\Gamma(1+x/2)^2 \Gamma(1- x) \Gamma(1 -\kappa x)  }\,.
 \end{equation} 
Notice this has simple poles at $x=2n$, which have residues that vanish for $\kappa$ half-integer.
Lastly we shall need\footnote{Because we are assuming that $\kappa$ is \textit{not} integer, $ \varrho(i \pm 0)$ is non-zero. If $\kappa<0$, then $ \varrho(i \pm 0)$ generically has a finite ambiguity.}
\begin{equation}
    \varrho(i \pm 0) =  8 e^{2b\mp i \pi \kappa} \frac{\Gamma(1+\kappa)}{\Gamma(1-\kappa)} = \frac{8}{\pi \kappa} e^{2b \mp i \pi \kappa} \Gamma(1+\kappa)^2 \sin(\pi \kappa)\,.
\end{equation}

Following again \cite{Marino:2021dzn},   the integral equation \eqref{eq:integraleqnu} is simplified by the introduction of   $P(\eta, v)$ given by
\begin{equation} \label{eq:Pdefn}
    e^{-2 B x}\delta\varrho(i x)  = - 2 iv e^{-\eta} P(\eta, v) \,,
\end{equation}
with a change of variables $(x, B) \rightarrow (\eta, v)$:
\begin{equation}
    \frac{1}{v } + a \log v = 2 B  \,, \quad x = v \eta \,.
\end{equation}
Here, $a$ is a constant determined by demanding that $P(\eta,v)$ is regular in $v$ with, in particular, no $\log(v)$ terms. From eq. \eqref{eq:deltavarrho}, we have that $ \delta \varrho(i  x )  \propto e^{ \tilde{a} x \log x} \sum d_n x^n$, where $\tilde{a}=-2 \kappa$, therefore this determines $\tilde{a}=a$. This yields an expansion of $P(\eta,v)$ given by
\begin{equation}
\begin{split}
    P(\eta,v) = d_{1,0} \eta + v \eta^2 (d_{2,0} + d_{2,1} \log(\eta)) + {\cal O}(v^2)\,, \\
    d_{1,0} = \pi \kappa\,, \quad d_{2,0}= 2 \pi \kappa ( 1 + (1- \gamma_E-\log(\kappa))\kappa - \log(2) )\,, \quad d_{2,1} = -2 \pi \kappa^2\,.
\end{split}
\end{equation}
With the introduction of an integral operator 
\begin{equation}
{\cal I}[f](\eta) =  - \frac{v}{\pi}\int_0^{\infty} d\eta' \frac{ e^{-\eta'} P(\eta',v) f(\eta')  }{\eta + \eta'}  \, , 
\end{equation}
 after this change of variables, eq. \eqref{eq:integraleqnu}  can be written as 
  \begin{equation} \label{eq:uetainteqn}
  u(\eta) =\mathfrak{u}(\eta)  + {\cal I}[u](\eta)   \, , 
  \end{equation} 
  in which the `seed' solution is given as 
\begin{equation} \label{eq:useed}
\mathfrak{u}(\eta)  =  \frac{1}{v\eta} +\frac{1}{v} \sum_n   \frac{ e^{-2 B v \eta_n}u_n \varrho_{n,\pm}  }{\eta_n + \eta  } \,.
\end{equation} 
The formal solution obtained by iteration is thus presented as  
\begin{equation} \label{eq:usolformal}
u(\eta) = \sum_{l=0}^{\infty} {\mathcal I}^l [ \mathfrak{u} ](\eta) \equiv   {\cal J}[\mathfrak{u} ](\eta) \, . 
\end{equation} 

To determine the unknown coefficients $u_n = u(\eta_n)$ we evaluate eq.\eqref{eq:uetainteqn} at $\eta = \eta_n = \mu n / v$ and define ${\cal I}_n[f] \equiv {\cal I}[f](\eta =\eta_n)$  to obtain
\begin{equation} \label{eq:uns}
u_n = \frac{1}{\mu n } + {\cal I}_n[u ]   + \frac{1}{\mu } \sum_m   \frac{ e^{-2 B v \eta_n} u_m \varrho_{m,\pm}  }{ m+ n   }\,.
\end{equation}  
Here we have made a slight adaptation compared to \cite{Marino:2021dzn} to suit the locations of the poles at   $x_n = \mu n$ (with $\mu=2$) (cf. the Gross-Neveu model for which $x_n = \frac{2 n +1}{\Upsilon} $ for some constant $\Upsilon$). To treat the exponentially small contributions coming from the residue term we introduce the small  parameter  
\begin{equation} \label{eq:qdefinB}
    q = e^{-2 B \mu} = e^{-\mu/v}v^{-\mu a}\,.
\end{equation}
Both the seed and formal solution, and the unkown values $u_n$, admit expansion in $q$  
\begin{equation}
\mathfrak{u} (\eta)  = \sum_{s=1} \mathfrak{u}^{(s)} (\eta)q^s\,, \quad  u(\eta)  = \sum u^{(s)} (\eta)q^s\,, \quad   u_n = \sum_{s=0} u_n^{(s)} q^s  \, .
\end{equation}

As the operator $\mathcal{J}$   does not introduce factors of $q$  we can construct the full solution order by order in $q$ noting $u^{(s)}(\eta) = {\cal J}[\mathfrak{u}^{(s)} ](\eta)$. Using \eqref{eq:useed} one finds that the first few terms\footnote{
For $n\geq 1$, we have in general
\begin{equation}
\mathfrak{u}^{(n)} (\eta) = \sum_{m=1}^n \frac{\varrho_{m,\pm} u_m^{(n-m)}}{v \eta + \mu m}\,.
\end{equation}} 
of the seed solution are given by 
\begin{equation} \label{eq:useedn}
\mathfrak{u}^{(0)}=\frac{1}{v\eta} \, , \quad 
\mathfrak{u}^{(1)} = \frac{ \varrho_{1,\pm} u_1^{(0)} }{v\eta +\mu }\, , \quad 
\mathfrak{u}^{(2)} = \frac{ \varrho_{1,\pm} u_1^{(1)} }{v\eta +\mu }+\frac{\varrho_{2,\pm} u_2^{(0)} }{v\eta +2 \mu }\, . 
\end{equation}

Applying the $q$-expansion to eq. \eqref{eq:uns}  we have that 
\begin{equation} 
u_n^{(0)}  = {\cal J}[\mathfrak{u}^{(0)} ] (\eta_n) = \frac{1}{\mu n } + {\cal I}_n[{\mathcal J}[\frac{1}{v\eta}  ]   ]  \,,
\quad 
u_n^{(1)}   =  {\cal I}_n[{\cal J}[\mathfrak{u}^{(1)}   ] ]   + \frac{1}{\mu }   \frac{ \varrho_{1,\pm}   u_1^{(0)}   }{ 1+ n   } \,.
\end{equation}  
Let us assume that $\varrho_{1, \pm} \neq 0$ (i.e. $\kappa$ is not half-integer), such that these two expressions are governing the leading behaviour. 
Suppose now we work formally\footnote{i.e. ignoring that $q$ is exponentially smaller than higher order polynomial terms in $v$.}  to leading order in $v$ and leading order in $q$ . Because each application of $\mathcal{I}$ carries a factor $v$, to leading order it is sufficient to consider only the identity operator  ${\cal J}=1+\dots $ which results in\footnote{The small $v$ limit can be taken also in the integral:
$$
\mathcal{I}\left[\frac{1}{v \eta}\right](\eta_n) = - \frac{v}{\pi} \int_0 ^ \infty \frac{e^{\eta'} d_{1,0} \eta }{v \eta' + n \mu} = - \frac{v}{\pi}\int_0 ^ \infty\left( \frac{e^{\eta'} d_{1,0}}{n \mu} + \mathcal{O}(v)\right) = - \frac{v d_{1,0}}{n \pi \mu} + \mathcal{O}(v^2)\,.
$$} 
  \begin{equation} 
u_n^{(0)}  =   \frac{1}{\mu n}- \frac{v}{n\pi \mu} d_{1,0} + {\cal O }(v^2) \,, \quad 
u_n^{(1)} = \frac{\varrho_{1,\pm} }{\mu^2(n+1)} - \frac{d_{1,0}\varrho_{1,\pm} }{\mu^2 \pi (n+1)} v + \mathcal{O}(v^2)\, . 
\end{equation} 
The leading orders of $u(\eta)$ are obtained by 
\begin{equation}
    \begin{aligned}
      u(\eta) =& \left[\mathfrak{u}^{(0)}  + \mathcal{I}[\mathfrak{u}^{(0)}]+ \mathcal{O}(v)\right] + q \left[\mathfrak{u}^{(1)} + \mathcal{I}[\mathfrak{u}^{(1)}] + \mathcal{O}(v^2)\right] + \mathcal{O}(q^2) \, .  
\end{aligned} 
\end{equation}

To implement the boundary condition that relates the chemical potential $h$ to $q,v,$ we will need
\begin{equation} \label{eq:uiinvandq}
    u(i) = u\left(\eta =\frac{1}{v}\right) = \left[  1 - \frac{d_{1,0}}{\pi} v + \mathcal{O}(v^2)\right] + \frac{q \varrho_{1,\pm} }{  \mu(1+\mu)}\left[ 1- \frac{d_{1,0} v}{\pi} + \mathcal{O}(v^2) \right] \, + \mathcal{O}(q^2)\,.
\end{equation}
The next step is to do the Legendre transform and calculate $\frac{e}{\rho^2}$ from $\Delta F$. This can then be used to compare against the perturbative calculation. The same procedure of resolving the cut away from the poles of $\rho$ and deforming the contour appropriately  yields 
\begin{equation} \label{eq:freeenergyinu2}
\begin{split}
      \Delta F (h) =&- \frac{h^2}{2\pi} u(i) G_+(0)^2 \bigg(  1 + \frac{v^2}{\pi} \int \frac{e^{-\eta} P(\eta,v)u(\eta)}{\eta v-1}  d\eta  \\&- e^{-2B}\varrho(i \pm \epsilon) u(i) - \sum_{n\geq1} \frac{q^n \varrho_{n, \pm} u_n}{\mu n-1} \bigg) \, . 
      \end{split}
\end{equation}
The leading orders of  eq. \eqref{eq:freeenergyinu2} are given by  
\begin{equation}
\begin{split}
      \Delta F (h) = &- \frac{G_+(0)^2 h}{2 \pi}\left( 1- \frac{2 d_{10}}{\pi}v + \mathcal{O}(v^2)\right) \times  \\
      &\left(1 - \rho(i \pm \epsilon) q^{1/\mu} + \frac{2 \rho_{1, \pm}}{\mu(1-\mu^2)}q - \frac{2 \rho_{1, \pm}\rho(i \pm \epsilon)}{\mu(1+\mu)}q^{1+1/\mu }+ \mathcal{O}(q^2) \right)  \, . 
\end{split}
\end{equation}
The first step of the Legendre transform is to relate $h$ to the parameters $q$ and $v$. This is done by substituting the expansion \eqref{eq:uiinvandq} for $u(i)$ into the boundary condition \eqref{eq:u(i)bc} which, for $\mu=2$, gives
\begin{equation}
    h = \frac{ m G_+(i)}{12 \pi G_+(0)} q^{-1/4}\big(\pi + d_{1,0} v+ \mathcal{O}(v^2)\big)\big(6- \rho_{1, \pm} q + \mathcal{O}(q^2)\big) \, . 
\end{equation}
As a consequence    $\rho = - \frac{d  \Delta F}{d h}$ is given by 
\begin{equation}\label{eq:rhoinvandq}
    \rho = \frac{G_+(i)G_+(0) m}{12 \pi^2}\Big(\pi - d_{1,0} v + \mathcal{O}(v^2)\Big)\Big(6 q^{-1/4} + \rho_{1,\pm} q^{3/4} + \mathcal{O}(q^{7/4})\Big) \,,
\end{equation}
from which  we obtain $\frac{e}{\rho^2}$ as a series in $v$ and $q$: 
\begin{equation} \label{eq:eoverrhoinqandv}
\begin{split}
    \frac{e}{\rho^2}   = \frac{1}{6 G_+(0)^2}\Big(\pi + 2 d_{1,0} v + \mathcal{O}(v^2)\Big)\Big( 3+ 3 \rho(i+ \pm \epsilon ) q^{1/2} + \rho_{1,\pm} q + \mathcal{O}(q^{3/2})\Big)\,.
\end{split}
\end{equation}

We will now write this expansion in terms of the coupling \eqref{eq:lambdacoupling} used in the previous Sections. Let us introduce a parameter exponentially small in $\gamma$, analogous to $q$ being exponentially small in $v$, given by $q_\gamma = e^{-4/\gamma}(\gamma/2)^{4 \kappa}$. We use \eqref{eq:lambdacoupling} to write $v$ as a series in $\gamma$ and $q_\gamma$. Substituting this series for $v$ (and $q=q(v)$) into \eqref{eq:eoverrhoinqandv}, we arrive at 
\begin{equation} \label{eq:IRtransseries}
\begin{split}
   \frac{8 \kappa}{\pi} \frac{e}{\rho^2} = &\Big( 1 + \kappa \gamma + \mathcal{O}(\gamma^2)\Big) -  8 e^{ \mp i \pi \kappa} \frac{\Gamma( \kappa)}{\Gamma(-\kappa)}q_\gamma^{1/2} ( 1+ \mathcal{O}(\gamma)) \\&+  2^{3-4\kappa} e^{ \mp 2 i \pi \kappa} \frac{\Gamma(2 \kappa)}{\Gamma(-2\kappa)}q_\gamma ( 1+ \mathcal{O}(\gamma))\,.
\end{split}
\end{equation}
We see that the first two coefficient of the perturbative series match precisely with eq. \eqref{eq:lambdapertser}. The presence of transseries parameters $q_\gamma = e^{-4/\gamma}(\gamma/2)^{4 \kappa}$ provides concrete predictions of the resurgent structure of the perturbative series. In particular, we compute the ambiguity of the transseries \eqref{eq:IRtransseries} due to the difference in result if the branch cut is left or right of the poles. To leading order in $q_\gamma$ and $\gamma$, it is given by
\begin{equation} \label{eq:transseriesambiguityIR}
    \frac{8 \kappa}{\pi}\left[ \left(\frac{e}{\rho^2}\right)_- - \left(\frac{e}{\rho^2}\right)_+ \right] = \frac{ 16 \pi i }{ \Gamma(-\kappa) \Gamma(1-\kappa)} \left(\frac{\gamma}{2}\right)^{2 \kappa} e^{-2/\gamma}\,.
\end{equation}
This is exactly the same ambiguity as obtained through the asymptotic analysis of our perturbative calculation - see eq.  \eqref{eq:IRambiguity}. We thus observe that the Borel-ambiguity of the perturbative series can be cancelled precisely by an ambiguity of a transmonomial. Therefore, the large order non-perturbative behaviour is unambiguous up to the order considered. This mirrors the fabled BZJJ ambiguity cancellation \cite{bogomolny1980calculation, zinn2004multi1, zinn2004multi2, ZinnJustin:1980uk} in a field theory context.

The analysis above only finds a source of the leading ambiguity on the positive real axis of the Borel plane. However, we can do a similar analysis to recover the Borel branch singularity on the negative real axis. The critical modification of the programme, as realised by \cite{Marino:2019eym}, is to deform the contour of the integral equation \eqref{eq:baseinteqnu} into the lower half plane, instead of the upper half plan. The critical analytic data is then given by the branch cut and residues at the negative imaginary axis. In the lower half plane, $\varrho(-ix)$ has residues at $x_n=2n +1$ and at $\tilde{x}_n := \frac{n}{\kappa} $. However, as the latter set of residues is unambiguous with respect to the branch cut, they do not contribute\footnote{They would be part of a transseries solution, but as they are unambiguous, they are not of interest to us currently. As a further side remark, when choosing $\kappa<0$, along the positive imaginary axis $\varrho(ix)$ also has such unambiguous residues at $x=\frac{n}{\kappa}$.}.  One subtlety when using this approach arises when computing $u(i)$. Deforming the contour of eq. \eqref{eq:baseinteqnu} to an envelopment of the negative imaginary axis picks up a residue at $\omega=-i$, which introduces a contribution of $u(-i) \rho(-i \pm \epsilon) q$ not present in the analysis above.  We will not present a detailed derivation as it is similar to the one above. Rather, we can report that the final result is a transseries with a leading ambiguity given by 
\begin{equation}
   \frac{8 \kappa}{\pi}\left[ \left(\frac{e}{\rho^2}\right)_- - \left(\frac{e}{\rho^2}\right)_+ \right]  = - \frac{ \pi i }{4 \Gamma(\kappa) \Gamma(1+\kappa)} \left(-\frac{\gamma}{2}\right)^{-2 \kappa} e^{2/\gamma}\,.
\end{equation}
This precisely matches the ambiguity of the perturbative sector around the negative real axis found in eq. \eqref{eq:UVambiguity}.
\section{Discussion} \label{sec:outlook}

In this note, we have studied the $\lambda$-model and brought it into the fold of resurgent analysis of \cite{volin2010mass, volin2011quantum, Marino:2019eym, Marino:2021dzn}. The model is particularly interesting, because, distinct from previously considered models, it has a interacting CFT fixed point in the UV. 

We have found a perturbative series for the energy density at finite chemical potential of the $\lambda$-model in Section \ref{sec:PertTBA} and identified with numerical techniques its asymptotic form in Section \ref{sec:Asymp}. A key feature is that the Borel resummation of the large order behaviour is ambiguous when taken along either the positive or negative real axis.   These ambiguities are exactly compensated/cancelled by a further ambiguity in a non-perturbative sector of a transseries solution in Section \ref{sec:transseries}.   These cancellations provide the $\lambda$-model with a robustly defined foundation which may serve as a paradigmatic example for other theories with asymptotic CFT behaviour.

Of particular note is that the leading ambiguity on the positive axis (and associated features in the Borel plane) vanishes for $\kappa \in \mathbb{Z}_{>0}$, i.e. when the WZW level $k$ divides the rank $N$ of the gauge group $SU(N)$.   This is reminiscent of   Cheshire-cat resurgence \cite{dorigoni2018grin, dorigoni2019picard, Kozcaz:2016wvy,Kamata:2021jrs} in which the full glory of resurgence only becomes apparent as you deform away from certain special points at which it truncates.

Let us finish with some broader questions to ponder following the analysis of the  $\lambda$-model that we hope might stimulate further investigations on the topic:

\begin{itemize}
 
    \item An interesting feature of the WZW CFT that defines the UV of the $\lambda$ model is that it exhibits level-rank duality  \cite{Naculich:1990hg}.  It would be valuable to understand the extent to which this property constrains, or is encapsulated, in the form of the transseries that defines the $\lambda$-model.   
     
    \item In a QFT it is sometimes possible to directly link poles/branch points in the Borel plane to finite action non-pertubative saddle configurations.  Remarkably, this can be done even in theories without instantons.  In a series of paper  \cite{cherman2015decoding, demulder2016resurgence, Schepers:2020ehn}  finite action `uniton' configurations of 1+1d integrable QFTs were matched to Borel poles of a quantum mechanics that followed by dimensional reduction with twisted boundary conditions (akin to a chemical potential as deployed here).   This poses a natural question: can  the features of the Borel plane we have found here via TBA methods be related to some finite action saddles?  Conversely, given the knowledge of such uniton configurations, what do they imply for the TBA method?  Achieving this would serve to put the semi-classical approaches of \cite{cherman2015decoding, demulder2016resurgence, Schepers:2020ehn} on a surer-footing in quantum field theory.

    \item On the other hand there are a class of ambiguities which don't (yet at least) have an interpretation as semi-classical saddles.  Instead they are renormalon ambiguities associated to certain classes of Feynman diagrams.   In \cite{DiPietro:2021yxb,Marino:2021six} it was shown how to construct such a series of diagrams which source the renormalon ambiguities in $1/N$ expansion of the $O(N)$ vector model, the Gross-Neveu and the $SU(N)$ PCM. It would be interesting to investigate if there are diagrams that are responsible for the ambiguities in the $\lambda$-models.
    
    \item The landscape of integrable models in two dimensions has been vastly expanded in recent years through variants of this $\lambda$-model, and the related Yang-Baxter $\sigma$-models.   It could be rewarding to deploy similar technique across this landscape included e.g. to models with multiple deformation parameters or theories based on  cosets rather than group manifolds. 
    
    \item In \cite{Appadu:2018ioy} the Quantum Inverse Scattering Method was applied to give a direct quantisation of the $\lambda$-models as a continuum limit of a spin $k$ Heisenberg spin-chain with inhomogeneities.  The parameter that governs the in-homogeneity becomes a mass.  Although the ground state of the system is quite a complicated Fermi sea, one can identify holes as certain particle excitations. After taking the continuum limit, one can obtain a TBA system for these excitations matching that of the QFT.   An exciting question is if the above resurgent structure can be given a similar {\em ab initio} derivation within the QISM framework.  
\end{itemize}

\section*{Acknowledgements}

DCT is supported by The Royal Society through a University Research FellowshipGeneralised Dualities in String Theory and Holography URF 150185 and in part by STFC grant ST/P00055X/1 as well as by the FWO-Vlaanderen through the project G006119N and Vrije Universiteit Brussel through the Strategic Research Program ``High-Energy Physics''.  LS is supported by a PhD studentship from The Royal Society and the grant RF\textbackslash ERE\textbackslash 210269. For the purpose of open access, the authors have applied a Creative Commons Attribution (CC BY) licence to any Author Accepted Manuscript version arising.  We thank M Mari\~no and T Reis for helpful comments on a draft and I Aniceto for comments relating to this project.


\renewbibmacro{in:}{}
\addcontentsline{toc}{section}{References}
\printbibliography

@article{Demulder:2015lva,
    author = "Demulder, Saskia and Sfetsos, Konstantinos and Thompson, Daniel C.",
    title = "{Integrable $\lambda$-deformations: Squashing Coset CFTs and $AdS_5\times S^5$}",
    eprint = "1504.02781",
    archivePrefix = "arXiv",
    primaryClass = "hep-th",
    doi = "10.1007/JHEP07(2015)019",
    journal = "JHEP",
    volume = "07",
    pages = "019",
    year = "2015"
}

@article{Borsato:2016ose,
    author = "Borsato, Riccardo and Wulff, Linus",
    title = "{Target space supergeometry of $\eta$ and $\lambda$-deformed strings}",
    eprint = "1608.03570",
    archivePrefix = "arXiv",
    primaryClass = "hep-th",
    reportNumber = "IMPERIAL-TP-LW-2016-03",
    doi = "10.1007/JHEP10(2016)045",
    journal = "JHEP",
    volume = "10",
    pages = "045",
    year = "2016"
}

@article{Itsios:2014vfa,
    author = "Itsios, Georgios and Sfetsos, Konstantinos and Siampos, Konstantinos and Torrielli, Alessandro",
    title = "{The classical Yang\textendash{}Baxter equation and the associated Yangian symmetry of gauged WZW-type theories}",
    eprint = "1409.0554",
    archivePrefix = "arXiv",
    primaryClass = "hep-th",
    reportNumber = "DMUS-MP-14-10",
    doi = "10.1016/j.nuclphysb.2014.10.004",
    journal = "Nucl. Phys. B",
    volume = "889",
    pages = "64--86",
    year = "2014"
}

@article{Georgiou:2019plp,
    author = "Georgiou, George and Sfetsos, Konstantinos and Siampos, Konstantinos",
    title = "{Strong integrability of $\lambda$-deformed models}",
    eprint = "1911.07859",
    archivePrefix = "arXiv",
    primaryClass = "hep-th",
    reportNumber = "CERN-TH-2019-186",
    doi = "10.1016/j.nuclphysb.2020.114923",
    journal = "Nucl. Phys. B",
    volume = "952",
    pages = "114923",
    year = "2020"
}

@article{Dunne:2015ywa,
    author = "Dunne, Gerald V. and Unsal, Mithat",
    title = "{Resurgence and Dynamics of O(N) and Grassmannian Sigma Models}",
    eprint = "1505.07803",
    archivePrefix = "arXiv",
    primaryClass = "hep-th",
    doi = "10.1007/JHEP09(2015)199",
    journal = "JHEP",
    volume = "09",
    pages = "199",
    year = "2015"
}

@article{Cherman:2013yfa,
    author = {Cherman, Aleksey and Dorigoni, Daniele and Dunne, Gerald V. and \"Unsal, Mithat},
    title = "{Resurgence in Quantum Field Theory: Nonperturbative Effects in the Principal Chiral Model}",
    eprint = "1308.0127",
    archivePrefix = "arXiv",
    primaryClass = "hep-th",
    reportNumber = "FTPI-MINN-13-28, UMN-TH-3218-23, DAMTP-2013-40",
    doi = "10.1103/PhysRevLett.112.021601",
    journal = "Phys. Rev. Lett.",
    volume = "112",
    pages = "021601",
    year = "2014"
}

@article{Kamata:2021jrs,
    author = {Kamata, Syo and Misumi, Tatsuhiro and Sueishi, Naohisa and \"Unsal, Mithat},
    title = "{Exact-WKB analysis for SUSY and quantum deformed potentials: Quantum mechanics with Grassmann fields and Wess-Zumino terms}",
    eprint = "2111.05922",
    archivePrefix = "arXiv",
    primaryClass = "hep-th",
    month = "11",
    year = "2021"
}

@article{Kozcaz:2016wvy,
    author = "Koz\c{c}az, Can and Sulejmanpasic, Tin and Tanizaki, Yuya and Ünsal, Mithat",
    title = "{Cheshire Cat resurgence, Self-resurgence and Quasi-Exact Solvable Systems}",
    eprint = "1609.06198",
    archivePrefix = "arXiv",
    primaryClass = "hep-th",
    reportNumber = "RBRC-1294",
    doi = "10.1007/s00220-018-3281-y",
    journal = "Commun. Math. Phys.",
    volume = "364",
    number = "3",
    pages = "835--878",
    year = "2018"
}

@article{Thompson:2019ipl,
    author = "Thompson, Daniel C.",
    editor = "Anagnostopoulos, Konstantinos and others",
    title = "{An Introduction to Generalised Dualities and their Applications to Holography and Integrability}",
    eprint = "1904.11561",
    archivePrefix = "arXiv",
    primaryClass = "hep-th",
    doi = "10.22323/1.347.0099",
    journal = "PoS",
    volume = "CORFU2018",
    pages = "099",
    year = "2019"
}

@article{ZinnJustin:1980uk,
    author = "Zinn-Justin, Jean",
    title = "{Perturbation Series at Large Orders in Quantum Mechanics and Field Theories: Application to the Problem of Resummation}",
    reportNumber = "FUB-HEP-8-80",
    doi = "10.1016/0370-1573(81)90016-8",
    journal = "Phys. Rept.",
    volume = "70",
    pages = "109",
    year = "1981"
}

@article{Hollowood:2015dpa,
    author = "Hollowood, Timothy J. and Miramontes, J. Luis and Schmidtt, David M.",
    title = "{S-Matrices and Quantum Group Symmetry of k-Deformed Sigma Models}",
    eprint = "1506.06601",
    archivePrefix = "arXiv",
    primaryClass = "hep-th",
    doi = "10.1088/1751-8113/49/46/465201",
    journal = "J. Phys. A",
    volume = "49",
    number = "46",
    pages = "465201",
    year = "2016"
}

@article{Appadu:2018ioy,
    author = "Appadu, Calan and Hollowood, Timothy J. and Price, Dafydd and Thompson, Daniel C.",
    title = "{Quantum Anisotropic Sigma and Lambda Models as Spin Chains}",
    eprint = "1802.06016",
    archivePrefix = "arXiv",
    primaryClass = "hep-th",
    doi = "10.1088/1751-8121/aadc6d",
    journal = "J. Phys. A",
    volume = "51",
    number = "40",
    pages = "405401",
    year = "2018"
}

@article{zamolodchikov1977exact,
  title={Exact two-particle S-matrix of quantum sine-Gordon solitons},
  author={Zamolodchikov, Al B},
  journal={Communications in Mathematical Physics},
  volume={55},
  number={2},
  pages={183--186},
  year={1977},
  publisher={Springer}
}

@article{Buscher:1987qj,
    author = "Buscher, T. H.",
    title = "{Path Integral Derivation of Quantum Duality in Nonlinear Sigma Models}",
    reportNumber = "ITP-SB-87-61",
    doi = "10.1016/0370-2693(88)90602-8",
    journal = "Phys. Lett. B",
    volume = "201",
    pages = "466--472",
    year = "1988"
}

@article{volin2010mass,
   author = "Volin, Dmytro",
    title = "{From the mass gap in $O(N)$ to the non-Borel-summability in $O(3)$ and $O(4)$ sigma-models}",
    eprint = "0904.2744",
    archivePrefix = "arXiv",
    primaryClass = "hep-th",
    doi = "10.1103/PhysRevD.81.105008",
    journal = "Phys. Rev. D",
    volume = "81",
    pages = "105008",
    year = "2010"
}

@article{volin2011quantum,
  title={Quantum integrability and functional equations: Applications to the spectral problem of AdS/CFT and two-dimensional sigma models},
  author={Volin, Dmytro},
  journal={Journal of Physics A: Mathematical and Theoretical},
  volume={44},
  number={12},
  pages={124003},
  year={2011},
  publisher={IOP Publishing},
  eprint = "1003.4725",
  archivePrefix = "arXiv",
  primaryClass = "hep-th"
}

@article{Sfetsos:2013wia,
    author = "Sfetsos, Konstadinos",
    title = "{Integrable interpolations: From exact CFTs to non-Abelian T-duals}",
    eprint = "1312.4560",
    archivePrefix = "arXiv",
    primaryClass = "hep-th",
    reportNumber = "DMUS-MP-13-23, DMUS--MP--13-23",
    doi = "10.1016/j.nuclphysb.2014.01.004",
    journal = "Nucl. Phys. B",
    volume = "880",
    pages = "225--246",
    year = "2014"
}

@article{sfetsos2015generalised,
    author = "Sfetsos, Konstantinos and Siampos, Konstantinos and Thompson, Daniel C.",
    title = "{Generalised integrable $\lambda$ - and $\eta$-deformations and their relation}",
    eprint = "1506.05784",
    archivePrefix = "arXiv",
    primaryClass = "hep-th",
    doi = "10.1016/j.nuclphysb.2015.08.015",
    journal = "Nucl. Phys. B",
    volume = "899",
    pages = "489--512",
    year = "2015"
}

@article{demulder2016resurgence,
    author = "Demulder, Saskia and Dorigoni, Daniele and Thompson, Daniel C.",
    title = "{Resurgence in $\eta$-deformed Principal Chiral Models}",
    eprint = "1604.07851",
    archivePrefix = "arXiv",
    primaryClass = "hep-th",
    doi = "10.1007/JHEP07(2016)088",
    journal = "JHEP",
    volume = "07",
    pages = "088",
    year = "2016"
}

@article{zinn2004multi1,
    author = "Zinn-Justin, J. and Jentschura, U.D.",
    title = "{Multi-instantons and exact results I: Conjectures, WKB expansions, and instanton interactions}",
    eprint = "quant-ph/0501136",
    archivePrefix = "arXiv",
    doi = "10.1016/j.aop.2004.04.004",
    journal = "Annals Phys.",
    volume = "313",
    pages = "197--267",
    year = "2004"
}

@article{zinn2004multi2,
    author = "Zinn-Justin, J. and Jentschura, U.D.",
    title = "{Multi-instantons and exact results II: Specific cases, higher-order effects, and numerical calculations}",
    eprint = "quant-ph/0501137",
    archivePrefix = "arXiv",
    doi = "10.1016/j.aop.2004.04.003",
    journal = "Annals Phys.",
    volume = "313",
    pages = "269--325",
    year = "2004"
}

@article{Hassan:1992gi,
    author = "Hassan, S. F. and Sen, Ashoke",
    title = "{Marginal deformations of WZNW and coset models from O(d,d) transformation}",
    eprint = "hep-th/9210121",
    archivePrefix = "arXiv",
    reportNumber = "TIFR-TH-92-61",
    doi = "10.1016/0550-3213(93)90429-S",
    journal = "Nucl. Phys. B",
    volume = "405",
    pages = "143--165",
    year = "1993"
}

@article{Gawedzki:1988hq,
    author = "Gawedzki, K. and Kupiainen, A.",
    title = "{G/h Conformal Field Theory from Gauged WZW Model}",
    reportNumber = "HU-TFT-88-29",
    doi = "10.1016/0370-2693(88)91081-7",
    journal = "Phys. Lett. B",
    volume = "215",
    pages = "119--123",
    year = "1988"
}

@article{Witten:1991mm,
    author = "Witten, Edward",
    title = "{On Holomorphic factorization of WZW and coset models}",
    reportNumber = "IASSNS-HEP-91-25",
    doi = "10.1007/BF02099196",
    journal = "Commun. Math. Phys.",
    volume = "144",
    pages = "189--212",
    year = "1992"
}

@article{Klimcik:2016rov,
    author = "Klim\v{c}\'ik, Ctirad",
    title = "{Poisson\textendash{}Lie T-duals of the bi-Yang\textendash{}Baxter models}",
    eprint = "1606.03016",
    archivePrefix = "arXiv",
    primaryClass = "hep-th",
    doi = "10.1016/j.physletb.2016.06.077",
    journal = "Phys. Lett. B",
    volume = "760",
    pages = "345--349",
    year = "2016"
}

@article{Klimcik:2015gba,
    author = "Klim\v{c}\'ik, Ctirad",
    title = "{\ensuremath{\eta} and \ensuremath{\lambda} deformations as E -models}",
    eprint = "1508.05832",
    archivePrefix = "arXiv",
    primaryClass = "hep-th",
    doi = "10.1016/j.nuclphysb.2015.09.011",
    journal = "Nucl. Phys. B",
    volume = "900",
    pages = "259--272",
    year = "2015"
}

@article{Hoare:2015gda,
    author = "Hoare, B. and Tseytlin, A. A.",
    title = "{On integrable deformations of superstring sigma models related to $AdS_n \times S^n$ supercosets}",
    eprint = "1504.07213",
    archivePrefix = "arXiv",
    primaryClass = "hep-th",
    reportNumber = "IMPERIAL-TP-AT-2015-02, HU-EP-15-21",
    doi = "10.1016/j.nuclphysb.2015.06.001",
    journal = "Nucl. Phys. B",
    volume = "897",
    pages = "448--478",
    year = "2015"
}

@article{Vicedo:2015pna,
    author = "Vicedo, Benoit",
    title = "{Deformed integrable \ensuremath{\sigma}-models, classical R-matrices and classical exchange algebra on Drinfel\textquoteright{}d doubles}",
    eprint = "1504.06303",
    archivePrefix = "arXiv",
    primaryClass = "hep-th",
    doi = "10.1088/1751-8113/48/35/355203",
    journal = "J. Phys. A",
    volume = "48",
    number = "35",
    pages = "355203",
    year = "2015"
}

@article{aniceto2018primer,
    author = "Aniceto, In{\^e}s and Ba\c{s}ar, Gok\c{c}e and Schiappa, Ricardo",
    title = "{A Primer on Resurgent Transseries and Their Asymptotics}",
    eprint = "1802.10441",
    archivePrefix = "arXiv",
    primaryClass = "hep-th",
    reportNumber = "NSF-ITP-17-153",
    doi = "10.1016/j.physrep.2019.02.003",
    journal = "Phys. Rept.",
    volume = "809",
    pages = "1--135",
    year = "2019"
}

@article{dunne2012resurgence,
  title={Resurgence and trans-series in Quantum Field Theory: the $\mathbb{CP}^{N-1} $ model},
  author={Dunne, Gerald V and {\"U}nsal, Mithat},
     eprint = "1210.2423",
    archivePrefix = "arXiv",
    primaryClass = "hep-th",
    doi = "10.1007/JHEP11(2012)170",
    journal = "JHEP",
    volume = "11",
    pages = "170",
    year = "2012"
}

@article{Hoare:2021dix,
    author = "Hoare, Ben",
    title = "{Integrable deformations of sigma models}",
    eprint = "2109.14284",
    archivePrefix = "arXiv",
    primaryClass = "hep-th",
    doi = "10.1088/1751-8121/ac4a1e",
    journal = "J. Phys. A",
    volume = "55",
    number = "9",
    pages = "093001",
    year = "2022"
}

@article{bogomolny1980calculation,
  title={Calculation of instanton-anti-instanton contributions in quantum mechanics},
  author={Bogomolny, EB},
  journal={Physics Letters B},
  volume={91},
  number={3-4},
  pages={431--435},
  year={1980},
  publisher={Elsevier}
}

@article{tHooft:1977xjm,
    author = "'t Hooft, Gerard",
    editor = "Zichichi, Antonino",
    title = "{Can We Make Sense Out of Quantum Chromodynamics?}",
    reportNumber = "PRINT-77-0723 (UTRECHT)",
    journal = "Subnucl. Ser.",
    volume = "15",
    pages = "943",
    year = "1979"
}

@article{Parisi:1978bj,
    author = "Parisi, G.",
    title = "{Singularities of the Borel Transform in Renormalizable Theories}",
    reportNumber = "LPTENS 78/8",
    doi = "10.1016/0370-2693(78)90101-6",
    journal = "Phys. Lett. B",
    volume = "76",
    pages = "65--66",
    year = "1978"
}

@article{Parisi:1978az,
    author = "Parisi, G.",
    title = "{On Infrared Divergences}",
    reportNumber = "LPTENS 78/11",
    doi = "10.1016/0550-3213(79)90298-0",
    journal = "Nucl. Phys. B",
    volume = "150",
    pages = "163--172",
    year = "1979"
}

@article{hollowood2014integrable,
    author = "Hollowood, Timothy J. and Miramontes, J.Luis and Schmidtt, David M.",
    title = "{An Integrable Deformation of the $AdS_5 \times S^5$ Superstring}",
    eprint = "1409.1538",
    archivePrefix = "arXiv",
    primaryClass = "hep-th",
    doi = "10.1088/1751-8113/47/49/495402",
    journal = "J. Phys. A",
    volume = "47",
    number = "49",
    pages = "495402",
    year = "2014"
}

@article{Sfetsos:2014cea,
    author = "Sfetsos, Konstantinos and Thompson, Daniel C.",
    title = "{Spacetimes for $\lambda$-deformations}",
    eprint = "1410.1886",
    archivePrefix = "arXiv",
    primaryClass = "hep-th",
    doi = "10.1007/JHEP12(2014)164",
    journal = "JHEP",
    volume = "12",
    pages = "164",
    year = "2014"
}

@article{Sfetsos:1994vz,
    author = "Sfetsos, Konstadinos",
    title = "{Gauged WZW models and nonAbelian duality}",
    eprint = "hep-th/9402031",
    archivePrefix = "arXiv",
    reportNumber = "THU-94-01",
    doi = "10.1103/PhysRevD.50.2784",
    journal = "Phys. Rev. D",
    volume = "50",
    pages = "2784--2798",
    year = "1994"
}

@article{Tseytlin:1993hm,
    author = "Tseytlin, Arkady A.",
    title = "{On A 'Universal' class of WZW type conformal models}",
    eprint = "hep-th/9311062",
    archivePrefix = "arXiv",
    reportNumber = "CERN-TH-7068-93",
    doi = "10.1016/0550-3213(94)90243-7",
    journal = "Nucl. Phys. B",
    volume = "418",
    pages = "173--194",
    year = "1994"
}

@article{Appadu:2015nfa,
    author = "Appadu, Calan and Hollowood, Timothy J.",
    title = "{Beta function of k deformed AdS$_{5}$ \texttimes{} S$^{5}$ string theory}",
    eprint = "1507.05420",
    archivePrefix = "arXiv",
    primaryClass = "hep-th",
    doi = "10.1007/JHEP11(2015)095",
    journal = "JHEP",
    volume = "11",
    pages = "095",
    year = "2015"
}

@article{Sfetsos:2014jfa,
    author = "Sfetsos, Konstadinos and Siampos, Konstadinos",
    title = "{Gauged WZW-type theories and the all-loop anisotropic non-Abelian Thirring model}",
    eprint = "1405.7803",
    archivePrefix = "arXiv",
    primaryClass = "hep-th",
    doi = "10.1016/j.nuclphysb.2014.06.012",
    journal = "Nucl. Phys. B",
    volume = "885",
    pages = "583--599",
    year = "2014"
}

@article{Evans:1994hi,
    author = "Evans, Jonathan M. and Hollowood, Timothy J.",
    title = "{Integrable theories that are asymptotically CFT}",
    eprint = "hep-th/9407113",
    archivePrefix = "arXiv",
    reportNumber = "CERN-TH-7293-94, SWAT-93-94-32",
    doi = "10.1016/0550-3213(94)00473-R",
    journal = "Nucl. Phys. B",
    volume = "438",
    pages = "469--490",
    year = "1995"
}

@article{ahn1990fractional,
  title={Fractional supersymmetries in perturbed coset CFTs and integrable soliton theory},
  author={Ahn, C and Bernard, D and LeClair, A},
  journal={Nuclear Physics B},
  volume={346},
  number={2-3},
  pages={409--439},
  year={1990},
  publisher={Elsevier}
}

@article{Balog:1993es,
    author = "Balog, J. and Forgacs, P. and Horvath, Z. and Palla, L.",
    editor = "Lust, D. and Weigt, G.",
    title = "{A New family of SU(2) symmetric integrable sigma models}",
    eprint = "hep-th/9307030",
    archivePrefix = "arXiv",
    reportNumber = "ITP-502-BUDAPEST",
    doi = "10.1016/0370-2693(94)90213-5",
    journal = "Phys. Lett. B",
    volume = "324",
    pages = "403--408",
    year = "1994"
}

@article{Hollowood:2014rla,
    author = "Hollowood, Timothy J. and Miramontes, J. Luis and Schmidtt, David M.",
    title = "{Integrable Deformations of Strings on Symmetric Spaces}",
    eprint = "1407.2840",
    archivePrefix = "arXiv",
    primaryClass = "hep-th",
    doi = "10.1007/JHEP11(2014)009",
    journal = "JHEP",
    volume = "11",
    pages = "009",
    year = "2014"
}

@incollection{witten1994non,
  title={Non-abelian bosonization in two dimensions},
  author={Witten, Edward},
  booktitle={Bosonization},
  pages={201--218},
  year={1994},
  publisher={World Scientific}
}

@article{dorigoni2019picard,
  title={Picard-Lefschetz decomposition and Cheshire Cat resurgence in 3D $\mathcal{N}=2$ field theories},
  author={Dorigoni, Daniele and Glass, Philip},
  journal={Journal of High Energy Physics},
  volume={2019},
  number={12},
  pages={1--40},
  year={2019},
  publisher={Springer}
}

@article{dorigoni2018grin,
  title={The grin of Cheshire cat resurgence from supersymmetric localization},
  author={Dorigoni, Daniele and Glass, Philip},
  journal={SciPost Physics},
  volume={4},
  number={2},
  pages={012},
  year={2018}
}

@article{cherman2015decoding,
    author = "Cherman, Aleksey and Dorigoni, Daniele and Unsal, Mithat",
    title = "{Decoding perturbation theory using resurgence: Stokes phenomena, new saddle points and Lefschetz thimbles}",
    eprint = "1403.1277",
    archivePrefix = "arXiv",
    primaryClass = "hep-th",
    reportNumber = "DAMTP-2014-17, UMN-TH-2239-14, FTPI-MINN-14-8",
    doi = "10.1007/JHEP10(2015)056",
    journal = "JHEP",
    volume = "10",
    pages = "056",
    year = "2015"
}

@article{Bajnok:2022xgx,
    author = "Bajnok, Zoltan and Balog, Janos and Vona, Istvan",
    title = "{The full analytic trans-series in integrable field theories}",
    eprint = "2212.09416",
    archivePrefix = "arXiv",
    primaryClass = "hep-th",
    month = "12",
    year = "2022"
}

@article{Zamolodchikov:1995xk,
    author = "Zamolodchikov, Alexei B.",
    title = "{Mass scale in the sine-Gordon model and its reductions}",
    doi = "10.1142/S0217751X9500053X",
    journal = "Int. J. Mod. Phys. A",
    volume = "10",
    pages = "1125--1150",
    year = "1995"
}

@article{Bajnok:2021dri,
    author = "Bajnok, Zolt\'an and Balog, J\'anos and Vona, Istv\'an",
    title = "{Analytic resurgence in the O(4) model}",
    eprint = "2111.15390",
    archivePrefix = "arXiv",
    primaryClass = "hep-th",
    doi = "10.1007/JHEP04(2022)043",
    journal = "JHEP",
    volume = "04",
    pages = "043",
    year = "2022"
}

@article{Abbott:2020qnl,
    author = "Abbott, Michael C. and Bajnok, Zolt\'an and Balog, J\'anos and Heged\'{u}s, \'Arp\'ad and Sadeghian, Saeedeh",
    title = "{Resurgence in the O(4) sigma model}",
    eprint = "2011.12254",
    archivePrefix = "arXiv",
    primaryClass = "hep-th",
    doi = "10.1007/JHEP05(2021)253",
    journal = "JHEP",
    volume = "05",
    pages = "253",
    year = "2021"
}

@article{Abbott:2020mba,
    author = "Abbott, Michael C. and Bajnok, Zolt\'an and Balog, J\'anos and Heged\'{u}s, \'Arp\'ad",
    title = "{From perturbative to non-perturbative in the O (4) sigma model}",
    eprint = "2011.09897",
    archivePrefix = "arXiv",
    primaryClass = "hep-th",
    doi = "10.1016/j.physletb.2021.136369",
    journal = "Phys. Lett. B",
    volume = "818",
    pages = "136369",
    year = "2021"
}

@article{Bajnok:2021zjm,
    author = "Bajnok, Zoltan and Balog, Janos and Hegedus, Arpad and Vona, Istvan",
    title = "{Instanton effects vs resurgence in the O(3) sigma model}",
    eprint = "2112.11741",
    archivePrefix = "arXiv",
    primaryClass = "hep-th",
    doi = "10.1016/j.physletb.2022.137073",
    journal = "Phys. Lett. B",
    volume = "829",
    pages = "137073",
    year = "2022"
}

@article{Bajnok:2022ucr,
    author = "Bajnok, Zoltan and Janik, Romuald A.",
    title = "{OPE coefficients and the mass-gap from the integrable scattering description of 2D CFT\textquoteright{}s}",
    eprint = "2209.10393",
    archivePrefix = "arXiv",
    primaryClass = "hep-th",
    doi = "10.1007/JHEP11(2022)128",
    journal = "JHEP",
    volume = "11",
    pages = "128",
    year = "2022"
}

@article{Bajnok:2022rtu,
    author = "Bajnok, Zoltan and Balog, Janos and Hegedus, Arpad and Vona, Istvan",
    title = "{Running coupling and non-perturbative corrections for O(N) free energy and for disk capacitor}",
    eprint = "2204.13365",
    archivePrefix = "arXiv",
    primaryClass = "hep-th",
    doi = "10.1007/JHEP09(2022)001",
    journal = "JHEP",
    volume = "09",
    pages = "001",
    year = "2022"
}

@article{Balog:1992cm,
    author = "Balog, J. and Naik, S. and Niedermayer, F. and Weisz, P.",
    title = "{Exact mass gap of the chiral SU(n) x SU(n) model}",
    reportNumber = "MPI-PH-92-22",
    doi = "10.1103/PhysRevLett.69.873",
    journal = "Phys. Rev. Lett.",
    volume = "69",
    pages = "873--876",
    year = "1992"
}

@article{DiPietro:2021yxb,
    author = "Di Pietro, Lorenzo and Mari\~no, Marcos and Sberveglieri, Giacomo and Serone, Marco",
    title = "{Resurgence and 1/N Expansion in Integrable Field Theories}",
    eprint = "2108.02647",
    archivePrefix = "arXiv",
    primaryClass = "hep-th",
    doi = "10.1007/JHEP10(2021)166",
    journal = "JHEP",
    volume = "10",
    pages = "166",
    year = "2021"
}

@article{Evans:1995dn,
    author = "Evans, Jonathan M. and Hollowood, Timothy J.",
    editor = "Mussardo, G. and Randjbar-Daemi, S. and Saleur, H.",
    title = "{Exact results for integrable asymptotically - free field theories}",
    eprint = "hep-th/9508141",
    archivePrefix = "arXiv",
    reportNumber = "CERN-TH-95-230, SWAT-94-95-63",
    doi = "10.1016/0920-5632(95)00622-2",
    journal = "Nucl. Phys. B Proc. Suppl.",
    volume = "45",
    number = "1",
    pages = "130--139",
    year = "1996"
}

@article{appadu2017quantum,
    author = "Appadu, Calan and Hollowood, Timothy J. and Price, Dafydd",
    title = "{Quantum Inverse Scattering and the Lambda Deformed Principal Chiral Model}",
    eprint = "1703.06699",
    archivePrefix = "arXiv",
    primaryClass = "hep-th",
    doi = "10.1088/1751-8121/aa7958",
    journal = "J. Phys. A",
    volume = "50",
    number = "30",
    pages = "305401",
    year = "2017"
}

@article{Marino:2019wra,
    author = "Mari\~no, Marcos and Reis, Tom\'as",
    title = "{Resurgence for superconductors}",
    eprint = "1905.09569",
    archivePrefix = "arXiv",
    primaryClass = "hep-th",
    doi = "10.1088/1742-5468/ab4802",
    month = "5",
    year = "2019"
}

@article{Marino:2019eym,
    author = "Mari\~no, Marcos and Reis, Tom\'as",
    title = "{Renormalons in integrable field theories}",
    eprint = "1909.12134",
    archivePrefix = "arXiv",
    primaryClass = "hep-th",
    doi = "10.1007/JHEP04(2020)160",
    journal = "JHEP",
    volume = "04",
    pages = "160",
    year = "2020"
}

@article{wiegmann1985exact,
  title={Exact solution of the $O(3)$ nonlinear two-dimensional sigma-model},
  author={Wiegmann, PB},
  journal={JETP Letters},
  volume={41},
  number={2},
  pages={95--100},
  year={1985}
}

@article{Marino:2021dzn,
    author = "Marino, Marcos and Miravitllas, Ramon and Reis, Tomas",
    title = "{New renormalons from analytic trans-series}",
    eprint = "2111.11951",
    archivePrefix = "arXiv",
    primaryClass = "hep-th",
    month = "11",
    year = "2021"
}

@article{Naculich:1990hg,
    author = "Naculich, Stephen G. and Schnitzer, Howard J.",
    title = "{Duality Between $SU(N)_k$ and $SU(k)_N$ {WZW} Models}",
    reportNumber = "BRX-TH-289",
    doi = "10.1016/0550-3213(90)90380-V",
    journal = "Nucl. Phys. B",
    volume = "347",
    pages = "687--742",
    year = "1990"
}

@article{Marino:2021six,
    author = "Marino, Marcos and Miravitllas, Ramon and Reis, Tomas",
    title = "{Testing the Bethe ansatz with large N renormalons}",
    eprint = "2102.03078",
    archivePrefix = "arXiv",
    primaryClass = "hep-th",
    doi = "10.1140/epjs/s11734-021-00252-4",
    journal = "Eur. Phys. J. ST",
    volume = "230",
    number = "12-13",
    pages = "2641--2666",
    year = "2021"
}

@article{polyakov1984goldstone,
  title={Goldstone fields in two dimensions with multivalued actions},
  author={Polyakov, Alexander M and Wiegmann, PB},
  journal={Physics Letters B},
  volume={141},
  number={3-4},
  pages={223--228},
  year={1984},
  publisher={Elsevier}
}

@article{hasenfratz1990exact1,
  title={The exact mass gap of the $O (3)$ and $O (4)$ non-linear $\sigma$-models in $d= 2$},
  author={Hasenfratz, Peter and Maggiore, Michele and Niedermayer, Ferenc},
  journal={Physics Letters B},
  volume={245},
  number={3-4},
  pages={522--528},
  year={1990},
  publisher={Elsevier}
}

@article{hasenfratz1990exact2,
  title={The exact mass gap of the $O(N)$ $\sigma$-model for arbitrary $N \geq 3$ in $d= 2$},
  author={Hasenfratz, Peter and Niedermayer, Ferenc},
  journal={Physics Letters B},
  volume={245},
  number={3-4},
  pages={529--532},
  year={1990},
  publisher={Elsevier}
}

@article{Forgacs:1991rs,
    author = "Forgacs, P. and Niedermayer, F. and Weisz, P.",
    title = "{The Exact mass gap of the Gross-Neveu model. 1. The Thermodynamic Bethe ansatz}",
    reportNumber = "MPI-PH-91-37",
    doi = "10.1016/0550-3213(91)90044-X",
    journal = "Nucl. Phys. B",
    volume = "367",
    pages = "123--143",
    year = "1991"
}

@article{Forgacs:1991ru,
    author = "Forgacs, P. and Niedermayer, F. and Weisz, P.",
    title = "{The Exact mass gap of the Gross-Neveu model. 2. The 1/N expansion}",
    reportNumber = "MPI-PH-91-38",
    doi = "10.1016/0550-3213(91)90045-Y",
    journal = "Nucl. Phys. B",
    volume = "367",
    pages = "144--157",
    year = "1991"
}

@article{Hollowood:1994np,
    author = "Hollowood, Timothy J.",
    title = "{The Exact mass gaps of the principal chiral models}",
    eprint = "hep-th/9402084",
    archivePrefix = "arXiv",
    reportNumber = "CERN-TH-7164-94, SWAT-93-94-26",
    doi = "10.1016/0370-2693(94)91089-8",
    journal = "Phys. Lett. B",
    volume = "329",
    pages = "450--456",
    year = "1994"
}

@article{Marino:2019fvu,
    author = "Mari\~no, Marcos and Reis, Tom\'as",
    title = "{A new renormalon in two dimensions}",
    eprint = "1912.06228",
    archivePrefix = "arXiv",
    primaryClass = "hep-th",
    doi = "10.1007/JHEP07(2020)216",
    journal = "JHEP",
    volume = "07",
    pages = "216",
    year = "2020"
}

@article{polyakov1983theory,
  title={Theory of nonabelian Goldstone bosons in two dimensions},
  author={Polyakov, A and Wiegmann, Paul B},
  journal={Physics Letters B},
  volume={131},
  number={1-3},
  pages={121--126},
  year={1983},
  publisher={Elsevier}
}

@article{Schepers:2020ehn,
    author = "Schepers, Lucas and Thompson, Daniel C.",
    title = "{Resurgence in the bi-Yang-Baxter model}",
    eprint = "2007.03683",
    archivePrefix = "arXiv",
    primaryClass = "hep-th",
    doi = "10.1016/j.nuclphysb.2021.115308",
    journal = "Nucl. Phys. B",
    volume = "964",
    pages = "115308",
    year = "2021"
}

\end{document}